\documentclass{aa}
\usepackage[varg]{txfonts}
\usepackage{natbib}
\bibpunct{(}{)}{;}{a}{}{,} 

\usepackage{amsmath}
\usepackage{graphicx}
\usepackage{graphics}
\usepackage{epsfig}
\usepackage{pdfpages} 
\usepackage{blindtext}
\usepackage{supertabular}
\usepackage{afterpage}
\usepackage{lscape}
\usepackage{tablefootnote}
\usepackage{subcaption}
\usepackage[flushleft]{threeparttable}
\usepackage{color, colortbl}
\usepackage[normalem]{ulem}
\usepackage{multirow}
\definecolor{lightgray}{gray}{0.9}
\usepackage{chemformula}

\usepackage{txfonts}

\usepackage{natbib,twoopt}
\usepackage[colorlinks=true,linkcolor=blue,citecolor=blue,urlcolor=blue,breaklinks=true]{hyperref} 
\bibpunct{(}{)}{;}{a}{}{,}             
\makeatletter
  \newcommandtwoopt{\citeads}[3][][]{\href{http://adsabs.harvard.edu/abs/#3}%
    {\def\hyper@linkstart##1##2{}%
     \let\hyper@linkend\@empty\citealp[#1][#2]{#3}}}
  \newcommandtwoopt{\citepads}[3][][]{\href{http://adsabs.harvard.edu/abs/#3}%
    {\def\hyper@linkstart##1##2{}%
     \let\hyper@linkend\@empty\citep[#1][#2]{#3}}}
  \newcommandtwoopt{\citetads}[3][][]{\href{http://adsabs.harvard.edu/abs/#3}%
    {\def\hyper@linkstart##1##2{}%
     \let\hyper@linkend\@empty\citet[#1][#2]{#3}}}
  \newcommandtwoopt{\citeyearads}[3][][]%
    {\href{http://adsabs.harvard.edu/abs/#3}
    {\def\hyper@linkstart##1##2{}%
     \let\hyper@linkend\@empty\citeyear[#1][#2]{#3}}}
\makeatother

\begin{document}
\title{Dust depletion of metals from local to distant galaxies II:} 
\subtitle{Cosmic dust-to-metal ratio and dust composition}

\author{Christina~Konstantopoulou\inst{\ref{inst-geneva}}\thanks{\email{christina.konstantopoulou@unige.ch}},
        Annalisa~De~Cia\inst{\ref{inst-geneva},\ref{inst-esogarch}},
        C\'{e}dric~Ledoux\inst{\ref{inst-eso}},
        Jens-Kristian~Krogager\inst{\ref{inst-lyon}},
        Lars~Mattsson\inst{\ref{inst-nordita}},
        Darach~Watson\inst{\ref{inst-dawn},\ref{inst-nbi}},
        Kasper~E.~Heintz\inst{\ref{inst-dawn},\ref{inst-nbi}},
        C\'{e}line~P\'{e}roux\inst{\ref{inst-esogarch},\ref{inst-LAM}},
        Pasquier~Noterdaeme\inst{\ref{inst-paris}},
        Anja~C.~Andersen\inst{\ref{inst-nbi2},\ref{inst-princeton}},
        Johan~P.~U.~Fynbo\inst{\ref{inst-dawn},\ref{inst-nbi}},
        Iris~Jermann\inst{\ref{inst-dawn},\ref{inst-dtu}},
        and Tanita~Ramburuth-Hurt\inst{\ref{inst-geneva}}}
        

\institute{Department of Astronomy, University of Geneva, Chemin Pegasi 51, 1290 Versoix, Switzerland \label{inst-geneva}
\and{European Southern Observatory, Karl-Schwarzschild Str. 2, 85748 Garching bei München, Germany \label{inst-esogarch}
\and European Southern Observatory, Alonso de C\'{o}rdova 3107, Vitacura, Casilla 19001, Santiago, Chile \label{inst-eso}
\and Centre de Recherche Astrophysique de Lyon, CNRS, Univ. Claude Bernard Lyon 1, 9 Av. Charles Andr\'{e}, 69230 Saint-Genis-Laval, France \label{inst-lyon}
\and Nordita, KTH Royal Institute of Technology and Stockholm University, Hannes Alfv\'{e}ns v\"ag 12, SE-106 91, Stockholm, Sweden\label{inst-nordita}
\and Cosmic Dawn Center (DAWN), Copenhagen, Denmark\label{inst-dawn}
\and Niels Bohr Institute, University of Copenhagen, Jagtvej 128, 2200 Copenhagen N, Denmark \label{inst-nbi}
\and Aix Marseille Universit\'{e}, CNRS, LAM (Laboratoire d’Astrophysique de Marseille) UMR 7326, 13388, Marseille, France  \label{inst-LAM}
\and Institut d'Astrophysique de Paris, CNRS-SU, UMR 7095, 98bis bd Arago, 75014, Paris, France \label{inst-paris}
\and Niels Bohr Institute, University of Copenhagen, \O{}ster Voldgade 5-7, 1353 Copenhagen K, Denmark\label{inst-nbi2}
\and Dept. of Astrophysical Sciences, Princeton University, Princeton, NJ 08544, USA\label{inst-princeton}
\and DTU Space, Technical University of Denmark, Elektrovej 327, 2800 Kgs.Lyngby, Denmark\label{inst-dtu}}}

\date{Received xxx; Accepted xxx}

 \abstract{The evolution of the cosmic dust content and the cycle between metals and dust in the interstellar medium (ISM) play a fundamental role in galaxy evolution. The chemical enrichment of the Universe can be traced through the evolution of the dust-to-metals ratio (DTM) and the dust-to-gas ratio (DTG) with metallicity. The physical processes through which dust is created and eventually destroyed still remain important open questions. We use a novel method to determine mass estimates of the DTM, DTG and dust composition in terms of fraction of dust mass contributed by element X (f$_{\rm{M}_{X}}$) based on our previous measurements of the depletion of metals in different environments (the Milky Way, the Magellanic Clouds, and damped Lyman-$\alpha$ absorbers, DLAs, toward quasars and towards gamma-ray bursts, GRBs), which were calculated from the relative abundances of metals in the ISM through absorption-line spectroscopy column densities observed mainly from VLT/UVES and X-shooter, and HST/STIS.
 We also derive the dust extinction from the estimated dust depletion ($A_{V, \rm depl}$) for GRB-DLAs, the Magellanic Clouds and the Milky Way and compare it with the $A_{V}$ estimated from extinction ($A_{V, \rm ext}$).
 We find that the DTM and DTG ratios increase with metallicity and with the dust tracer [Zn/Fe]. This suggests that grain growth in the ISM is a dominant process of dust production, at least in the metallicity range (-2 $\leq$ [M/H]$_{\rm{tot}}$ $\leq$ 0.5) and redshift range ($0.6 < z < 6.3$) that we are studying. The increasing trend of the DTM and DTG with metallicity is in good agreement with a dust production and evolution hydrodynamical model. Our data suggest that the stellar dust yield is much lower (about 1$\%$) than the metal yield and thus that the overall amount of dust in the warm neutral medium that is produced by stars is much lower, than previously estimated. The global neutral gas metallicity is decreasing over cosmic time and is traced similarly by quasar- and GRB-DLAs. We find that $A_{V,\rm depl}$ is overall lower than $A_{V, \rm ext}$ for the Milky Way and a few Magellanic Clouds lines of sight, a discrepancy that is likely related to the presence of carbonaceous dust associated with dense clumps of cold neutral gas. For the other environments, we find an overall good agreement between the $A_{V, \rm ext}$ and $A_{V, \rm depl}$. We show that the main elements (f$_{\rm{M}_{X}}$ > 1$\%$) that contribute to the dust composition are, by mass, O, Fe, Si, Mg, C, S, Ni and Al for all the environments, with Si, Mg and C being equivalent contributors. There are, nevertheless, variations in the dust composition depending on the overall amount of dust. The abundances measured at low dust regimes in quasar- and GRB-DLAs suggest the presence of pyroxene and metallic iron in dust. These results give important information on the dust and metal content of galaxies across cosmic times, from the Milky Way up to $z = 6.3$.
 }

\authorrunning{C. Konstantopoulou et al.}
\titlerunning{Dust depletion of metals from local to distant galaxies II}
\maketitle

\section{Introduction}
The interplay between the production of metals by stars and the formation and evolution of cosmic dust plays a fundamental role in the chemical enrichment of the interstellar medium (ISM). The cycle of exchanging material between stars, interstellar gas and cosmic dust drives galactic evolution \citep{Maiolino2019}. Yet the direct observations of the chemical properties of the baryon cycle remain a big challenge.

Metals are formed in stars and are ejected into the ISM by strong stellar winds and supernova explosions. More than 90$\%$ of the baryons are found in the gas phase at $z$ > 2 \citep{Peroux2020}. These metals in the ISM can be incorporated into the next generation of stars. While substantial amount of metals are found in neutral gas, large fractions of these metals are instead locked into dust grains, an effect called dust depletion \citep{Field1974, Savage1996, Phillips1982,Savaglio2003, Jenkins2009, DeCia2016, Roman-Duval2021}. Potential channels for the production of dust are the cooling ejecta of supernovae \citep{Dunne2003,Matsuura2015}, grain growth in the ISM \citep{Draine2003,Mattsson2012,Dwek2016,DeCia2016} and asymptotic giant branch (AGB) stars \citep{Gail2009, Hofner2018}. Around 30$\%$ of the light in the Universe is re-radiated by dust grains in the infrared, altering the light that we observe \citep{Calzetti1995,Witt2000}. Dust is also necessary in the process of star formation, because it is an important direct coolant for star formation and acts as a catalyst for the production of molecular hydrogen \citep{Hollenbach1971ApJ, Hollenbach1979}. Therefore, characterizing the dust properties and also the cycle of metals between the dust and gas is of fundamental importance in our understanding of galaxy formation and evolution.

The fraction of metals in dust can be described by the dust-to-metal ratio, and another indicator of the dust content is the dust-to-gas ratio \citep{Draine2003}. These properties can help us understand the production and destruction mechanisms of the dust and its evolution. They can be used to estimate a number of dust properties, including the dust surface density \citep[see][]{Peroux2023} and the dust composition. One way to infer the dust composition is to observe the abundances of metals in the gas phase ISM through absorption-line spectroscopy and taking into account their depletion into dust grains \citep[e.g,][]{Roman-Duval2022a}. In \citet{Konstantopoulou2022} (hereafter Paper I), we characterized the dust depletion of 18 metals in the Milky Way, the Magellanic Clouds, QSO- and GRB-DLAs using the ISM relative abundances of metals, following the prescription of \citep{DeCia2016}. Here we use the depletion estimates from Paper I to infer the dust properties, namely the dust-to-gas (DTG) and dust-to-metal (DTM) ratios, dust composition, and dust extinction in a wide range of galactic environments (the Milky Way, the Magellanic Clouds, QSO- and GRB-DLAs) and study how they evolve over cosmic time.

The paper is organized as follows. In Section 2 we briefly present the samples and the method that is used for the dust depletion estimates. In Section 3 we infer the DTM, DTG, the extinction and dust composition from depletion. We discuss our results in Section 4, and finally summarize and conclude in Section 5.

Throughout the paper we use a linear unit for the column densities N in terms of ions cm$^{-2}$. We refer to relative abundances of elements X and Y as $[\rm X/Y] \equiv$ log$\frac{N(\rm{X})}{N(\rm Y)} - $ log$\frac{N(\rm X)_{\odot}}{N(\rm Y)_{\odot}}$, where reference solar abundances are taken from \citet{Asplund2021} following the recommendations of \citet{Lodders2009} (see Table 1 in Paper I). We adopt a $\Lambda$ cold dark matter cosmology ($\Lambda$CDM) with H$_{0}$ = 67.7\,km\,s$^{-1}$\,Mpc$^{-1}$,  $\Omega_{ \rm M}$ = 0.3 and $\Omega_{\Lambda}$ = 0.7 \citep{Planck2020}.

\section{Data Samples}
\label{sec:sample}
In this section we briefly present the observational sample that we use for our analysis and the method that is used to estimate the dust depletion of all the metals for the Milky Way, the Magellanic Clouds and for high-redshift QSO- and GRB-DLAs. These can be found in more detail in Paper I.

In Paper I we used a large compilation of metal column densities, measured in a consistent way, in the neutral ISM for the Milky Way, the Magellanic Clouds, QSO- and GRB-DLAs to characterize the dust depletion of 18 metals (C, P, O, Cl, Kr, S, Ge, Mg, Si, Cu, Co, Mn, Cr, Ni, Al, Ti, Zn, and Fe). Our sample probes gas in a wide range of galaxy types and regions. QSO-DLAs trace gas-rich regions in the outskirts of galaxies \citep{Prochaska2007, Fynbo2008}, while GRB-DLAs probe the inner regions of galaxies hosting GRBs in a way more similar to the Milky Way and the Magellanic Clouds samples but at high redshifts.

The Milky Way sample is from \citet{Jenkins2009}, \citet{DeCia2021}, \citet{Welty2010}, and \citet{Phillips1982}. The Large Magellanic Cloud (LMC) sample from \citet{Roman-Duval2021}, the Small Magellanic Cloud (SMC) from \citet{Welty2010}, \citet{Tchernyshyov2015}, \citet{Jenkins2017} and the QSO-DLAs sample is from \citet{DeCia2016}, \citet{Berg2015}, \citet{DeCia2018}.
We also use 36 GRB-DLAs from \citet{Savaglio2003, Shin2006,Prochaska2007, Piranomonte08, Ledoux2009,DElia11,Wiseman2017, Zafar2019,Bolmer2019, Saccardi2023}, as compiled in \citet{Heintz2023}, who also presents three new GRB-DLAs.

The column densities are measured mainly from spectra obtained with UVES/VLT, X-Shooter/VLT and STIS/HST and were homogenized to the newest oscillator strengths, listed in Paper I.
The details of the full sample properties are presented in Paper I. 

\section{Methods and results}
\label{sec:methres}

\subsection{Dust depletion measurements}
\label{sec:dustdepl}

In this work we adopt the dust depletions of each metal from Paper I, which were estimated from the relative abundances of metals \citep{DeCia2016}, based on the fact that different metals have different tendencies of being incorporated into dust grains. We note that this is a different approach from the one of \citet{Roman-Duval2021}, who also derived dust properties based on depletions, but estimated the depletions from the observed abundances and assuming a given metallicity of the gas. Here we briefly present our method.

The depletion of an element X ($\delta_{\rm{X}}$) can be estimated from the relative abundances of two elements that have different refractory properties but follow each other nucleosynthetically. If X is a refractory element and Y is volatile (i.e., is not easily depleted into dust, like Zn) then the dust depletion of X can be expressed using the ratio of the two elements [X/Y]. {\citet{DeCia2016} and Paper I relied on [X/Zn] and their relation to the dust tracer to estimate the depletion of X. To trace the overall amount of dust we used the dust tracer [Zn/Fe]. Two corrections are applied: first the dependence on the depletion of Zn is removed by assuming $\delta_{\rm{Zn}}$ = -0.27$\times$[Zn/Fe], which is derived from the Milky Way and QSO-DLAs \citep{DeCia2016} and second a correction is applied to account for nucleosynthesis effects, such as $\alpha$-element enhancement and Mn underabundance, as described in Paper I.

The depletion of X correlates linearly with the dust tracer, such as [Zn/Fe], but others, such as [Si/Ti] or [O/Si] are possible and thus the dust depletion can then be expressed by a simple linear relation as,
\begin{equation}
    \delta_X = A2_X +B2_X \times \mathrm{[Zn/Fe]},
    \label{deltaxcoeff}
\end{equation}
where the A2$_{X}$ and B2$_{X}$ coefficients for all the metals are the results of the linear fits to the data and are presented in Table 4 of Paper I. These results are consistent with \citet{DeCia2016}. In Paper I the coefficients are estimated using two different assumptions on $\alpha$-element enhancement and Mn underabundance in the Magellanic Clouds. Here we adopt the coefficients that were estimated using the assumption of constant $\alpha$-element enhancement and Mn underabundance and are reported in Table \ref{coefficients} for completeness.

\begin{table}[!t]
\caption{Coefficients A2$_{X}$ and B2$_{X}$ resulting from the linear fit $\delta_{\rm{X}}$ = A2$_{X}$ +B2$_{X}$ $\times$  $\mathrm{[Zn/Fe]}$ of the depletion sequences of metals presented in Paper I and P from \citet{Konstantopoulou2023}.}
\label{coefficients}
\centering
\begin{tabular}{c c c c} 
\hline\hline 
Element X & A2$_{x}$ & B2$_{x}$\\
\hline
C & 0.00 & $-$0.10$\pm$0.10 \\
P & 0.08$\pm$0.05 & $-$0.26$\pm$0.08\\
O & 0.00  & $-$0.20$\pm$0.05\\
Cl & 0.00 & $-$0.12$\pm$0.09\\
Kr & 0.00  & $-$0.04$\pm$0.09\\
S & 0.01$\pm$0.02  & $-$0.48$\pm$0.04\\
Zn & 0.00$\pm$0.01 & $-$0.27$\pm$0.03\\
Ge & 0.00 & $-$0.40$\pm$0.04\\
Mg & 0.01$\pm$0.03 & $-$0.66$\pm$0.04\\
Si & $-$0.04$\pm$0.02 & $-$0.75$\pm$0.03\\
Cu & 0.00 & $-$0.73$\pm$0.04\\
Co & 0.00 & $-$0.89$\pm$0.19\\
Mn & 0.07$\pm$0.02 & $-$1.03$\pm$0.03\\
Cr & 0.12$\pm$0.01 & $-$1.30$\pm$0.01\\
Ni & 0.07$\pm$0.02   & $-$1.31$\pm$0.03\\
Fe &$-$0.01$\pm$0.03  & $-$1.26$\pm$0.04\\
Al & 0.00   & $-$1.66$\pm$0.35\\
Ti & $-$0.07$\pm$0.03 & $-$1.67$\pm$0.04\\ \hline
\end{tabular}
\end{table}


\subsection{Dust-to-metal and dust-to-gas mass ratios}
\label{sec:dtmdtg}

The total abundance of element X can be expressed as:

\begin{equation}
    \left(\frac{N(\rm X)}{\textit N(\rm H)}\right)_{\rm{tot}} = 10^{([\rm X/H]_\odot+[\textit M/\rm H]_{\rm{tot}})},
\end{equation}
where [X/H]$_{\odot}$ is the solar abundance of element X and [\textit{M}/$\rm{H}$]$_{\rm{tot}}$ is the total dust-corrected metallicity of the gas.

Then the DTM mass ratio can be calculated including all the metals as:

\begin{equation}
\begin{aligned}
    \rm{DTM} = \frac{M_{dust}}{M_{metals}} = \frac{\Sigma_{\rm X_i}\, (1-10^{\delta_{X_i}})~10^{[\rm X_i/H]_{\odot}}~\rm W_{X_i}}{\Sigma_{\rm X_i}\, 10^{[\rm X_i/H]_{\odot}}~\rm W_{X_i}},
\end{aligned}
\label{dtm}
\end{equation}
where $\delta_{X}$ is the dust depletion of element X, which are taken from Paper I and W$_{X}$ the atomic weight of element X. The total dust-corrected metallicity of the gas [\textit{M}/$\rm{H}$]$_{\rm{tot}}$ cancels out and thus the dust-to-metal ratio is estimated independently of the total metallicity. \footnote{We note that Eq. \ref{dtm} is equivalent to eq. 9 of \citet{Roman-Duval2022a}, but the depletions are estimated from the observed abundances. In \citet{DeCia2016} the calculation of DTM is based on ($1-10^{\delta_{X}}$) and only for Fe.} The sum is over all metals that have an elemental abundance $12 + \rm log\,(X/H) > 3$. In total we use 29 metals, namely, C, P, O, Cl, Kr, S, Ge, Mg, Si, Cu, Co, Mn, Cr, Ni, Al, Ti, N, Ne, Ar, F, Na, K, Ga, V, Sc, Se, Li, Zn, and Fe. The abundance measurements of these metals are done in the neutral gas of the ISM in their dominant ionization state, which is mostly singly ionized. In some cases, some elements should be volatile, although their depletion is not assessed in Paper I. In such cases (namely, N, Ne, Ar, F, Na, K, Ga, V, Sc, Se and Li) we assume that $\delta_{X}$ = 0. These metals make a small difference in the calculation of the DTM and DTG, but we include them because they contribute to the mass of the metals by summing their atomic weights in the denominator of Eq. \ref{dtm}.
The calculation of the depletions using the method of relative abundances is based on the availability of Zn and Fe. Thus, when Zn and/or Fe were not available, no estimate of the depletion was made in Paper I. In case $\delta_{X}$ was not estimated in Paper I, but [Zn/Fe] has been observed, we use the coefficients A2$_{X}$ and B2$_{X}$ listed in Table \ref{coefficients} to estimate the depletion $\delta_{X}$ using Eq. \ref{deltaxcoeff}. 
The dust depletion is typically defined as a negative number, because it removes metals from the gas-phase. In a few cases, however, the observed [Zn/Fe] is slightly negative, and typically consistent with zero within the uncertainties. In such cases, the values of depletion ($\delta_{X}$) that we strictly derive from the [Zn/Fe] would be slightly positive, which is non-physical (i.e. the presence of dust removes metals from the gas phase, does not add them). We therefore assign zero depletion in such cases. For the cases of GRB-DLAs with no observed [Zn/Fe], we use the estimated [Zn/Fe]$_{\rm fit}$ from \citet{Heintz2023} to calculate the expected depletions from the coefficients A2$_{X}$ and B2$_{X}$, listed in Table \ref{coefficients} and using Eq. \ref{deltaxcoeff}. [Zn/Fe]$_{\rm fit}$ is equivalent to the observed [Zn/Fe], but is derived from the abundances of all the available metals. 



The DTG is the total dust mass relative to the total gas mass. It can be expressed as:

\begin{equation}
    \rm{DTG} = \frac{M_{dust}}{M_{gas}} = DTM \times 10^{[M/H]_{\rm{tot}}} \times Z_{\odot},
\label{dtg}
\end{equation}
where DTM is defined in Eq. \ref{dtm}, [\textit{M}/$\rm{H}$]$_{\rm{tot}}$ is the total dust-corrected metallicity of the gas and Z$_{\odot}$ = 0.0139 is the solar metallicity taken from \citet{Asplund2021}. 
Figures \ref{fig:dtm_dtg_znfe} and \ref{fig:dtm_dtg_mh} show the DTM (top panel) and the DTG (bottom panel) with respect to the dust tracer [Zn/Fe] and the total dust-corrected metallicity [\textit{M}/$\rm{H}$]$_{\rm{tot}}$ respectfully and for all the environments. Our measurements are given in Tables \ref{dlastable}- \ref{grbtable}.

\begin{figure*}
    \centering
    \includegraphics[width=0.6\textwidth]{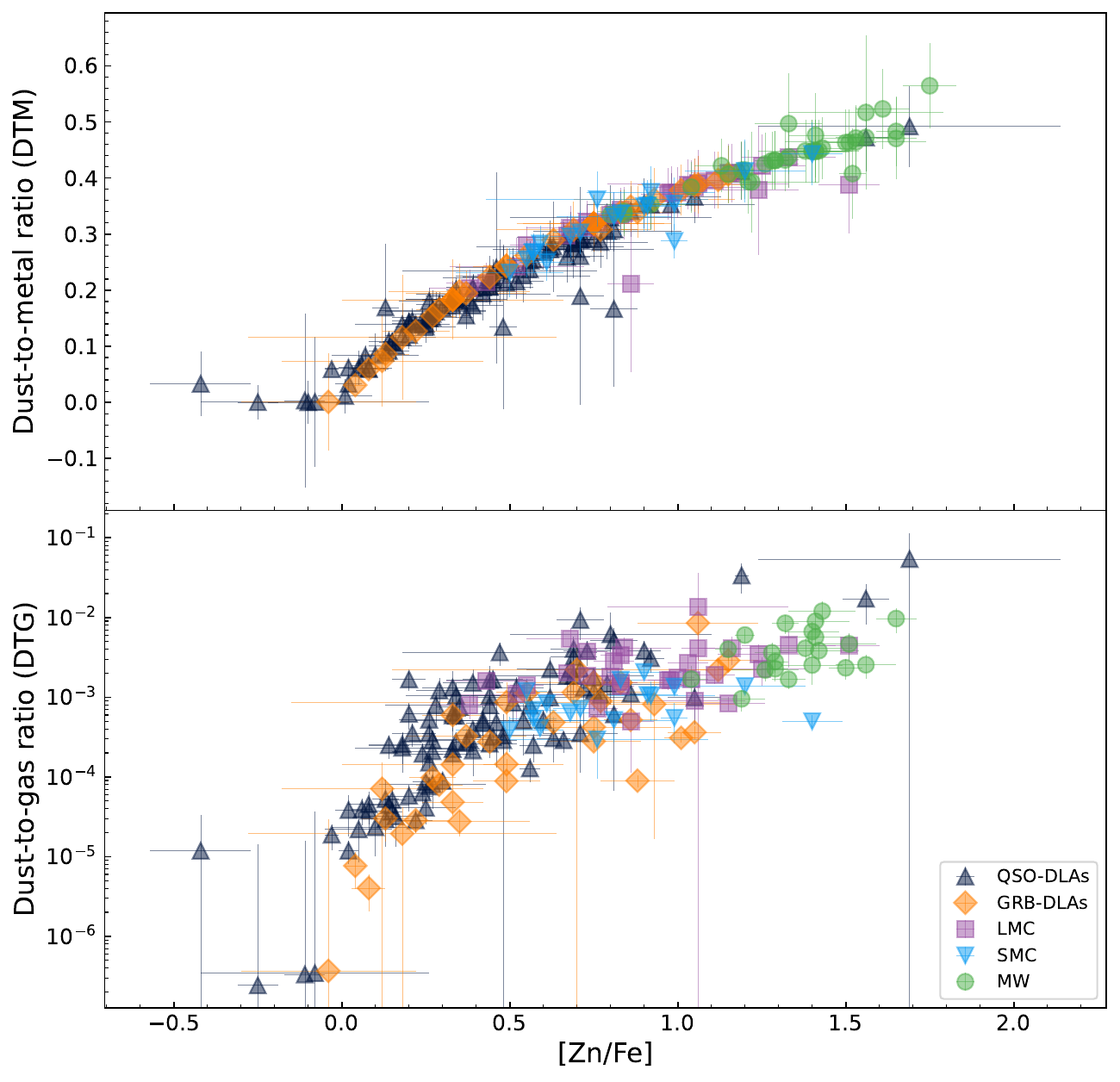}
    \caption{(Top): Dust-to-metal ratio as a function of the dust tracer [Zn/Fe]. (Bottom): Dust-to-gas ratio as a function of [Zn/Fe]. The black triangles are for QSO-DLAs, the purple squares are for the LMC, the blue triangles are for the SMC, the orange diamonds for the GRB-DLAs and the green circles are for the Milky Way.}
    \label{fig:dtm_dtg_znfe}
\end{figure*}

\begin{figure*}
    \centering
    \includegraphics[width=0.6\textwidth]{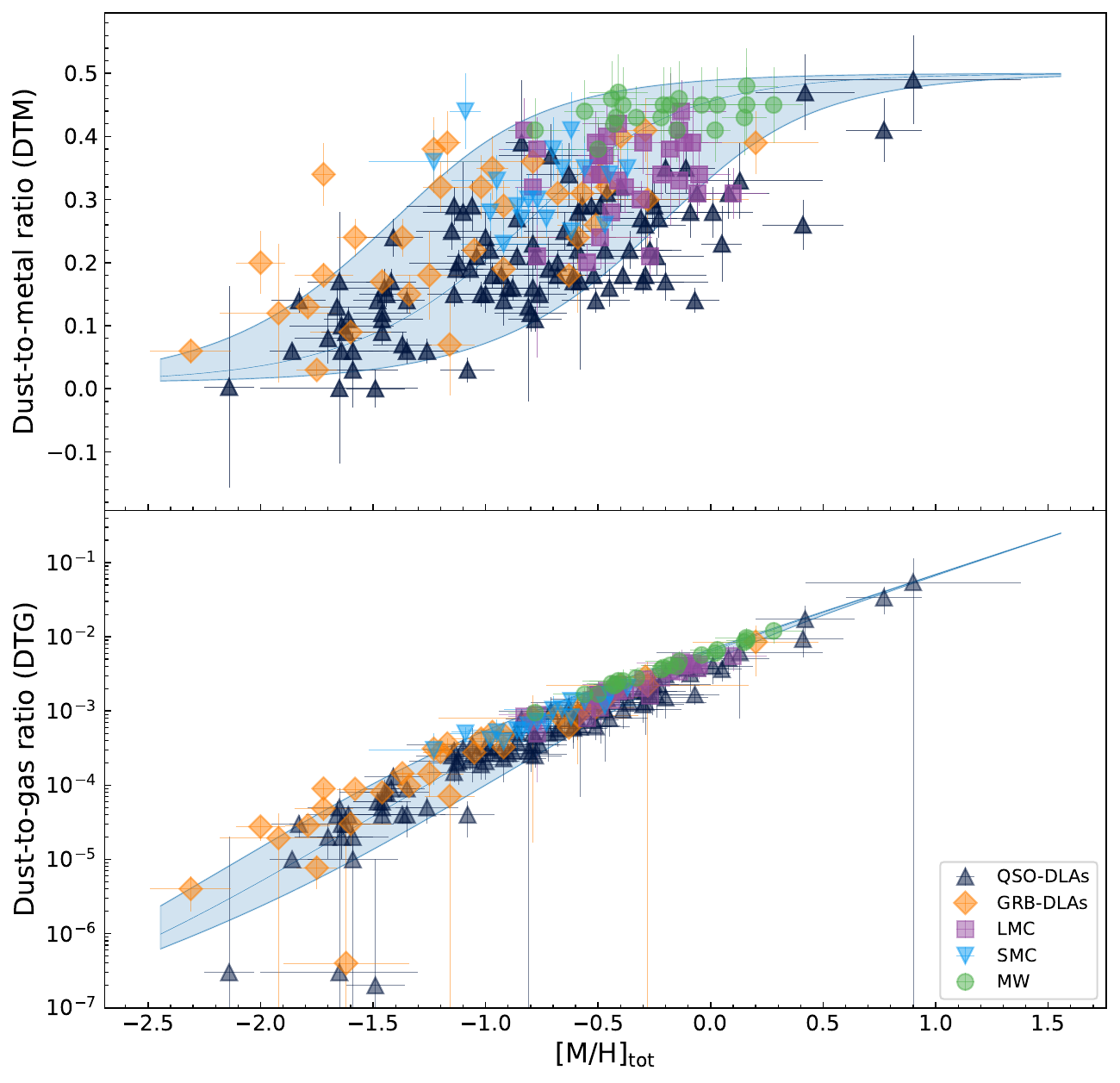}
    \caption{(Top): Dust-to-metal ratio as a function of the total dust-corrected gas metallicity [\textit{M}/$\rm{H}$]$_{\rm{tot}}$. (Bottom): Dust-to-gas ratio as a function of the total dust-corrected gas metallicity [\textit{M}/$\rm{H}$]$_{\rm{tot}}$. The symbols are the same as in Fig. \ref{fig:dtm_dtg_znfe}. For the Milky Way metallicities $[M/\rm H]_{\rm{tot}}$ are from \citet{DeCia2021}. There is a clear increasing trend of the DTM and DTG ratios with [\textit{M}/$\rm{H}$]$_{\rm{tot}}$ and a tight correlation of the DTG with [\textit{M}/$\rm{H}$]$_{\rm{tot}}$. The blue shaded regions show the analytic the dust production and evolution model from \citet{Mattsson2020} with varying $\delta$ = $\epsilon$ from 1 to 16.}
    \label{fig:dtm_dtg_mh}
\end{figure*}


\subsection{Dust composition}
\label{sec:dustcomp}

We estimate the fraction of dust mass contributed by element X, f$_{\rm{M}_{X}}$ as,

\begin{equation}
\begin{aligned}
    f_{\rm M_\textit X} = \frac{(1-10^{\delta_X})~10^{[\rm X/H]_\odot}~\rm W_X}{\Sigma_{X_i}\, (1-10^{\delta_{X_i}})~10^{[\rm X_{\textit i}/\rm H]_\odot}~\rm W_{X_i}},
\label{MFDx}
\end{aligned}
\end{equation}
where [X/H]$_{\odot}$ is the solar abundance of element X, $\delta_{X}$ is the dust depletion of element X and W$_{X}$ the atomic weight of element X.\footnote{We note that Eq. \ref{MFDx} is equivalent to Eq. 12 of \citet{Roman-Duval2022a}, but the depletions are estimated in a different way (see Sect. \ref{sec:dustdepl}).}

\begin{figure*}[h!]
    \includegraphics[width=0.74\textwidth]
    {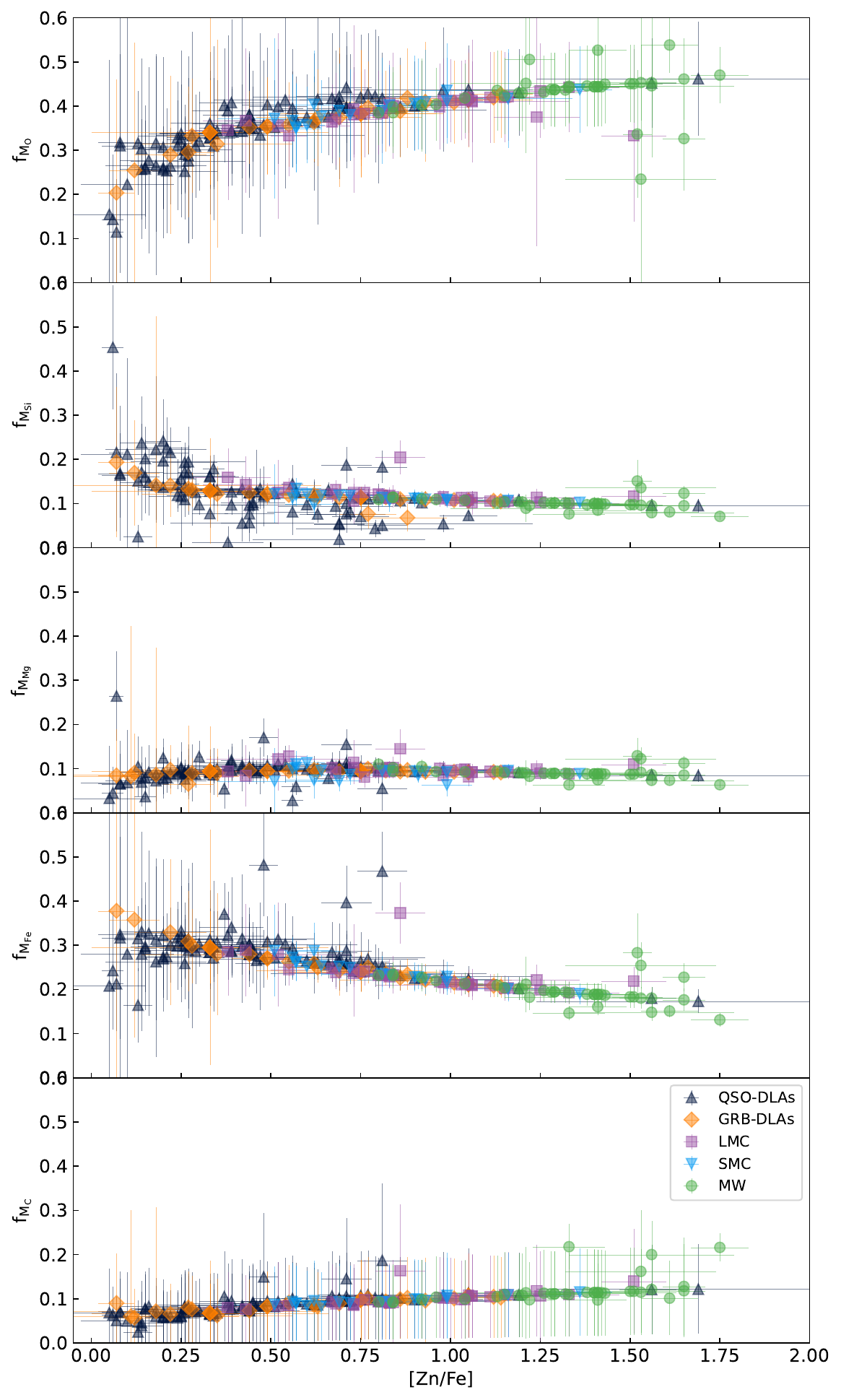}
    \centering
    \caption{Fraction of dust mass contributed by element X, f$_{\rm{M}_{X}}$ as a function of the dust tracer [Zn/Fe] for the five most abundant contributors in dust (O, Fe, Si, Mg, C) for QSO- and GRB-DLAs, LMC, SMC and the Milky Way.}
    \label{fig:dust_composition}
\end{figure*}

Figure \ref{fig:dust_composition} shows the evolution of dust composition with [Zn/Fe] for the elements that contribute at least 1$\%$ to the dust composition (f$_{\rm{M}_{X}}$ > 1$\%$). In case a measurement of $\delta_{X}$ is unavailable for a given system, we calculate it from the measurements of dust depletion of a given metal. This is done using Eq. \ref{deltaxcoeff} and the coefficients presented in Table \ref{coefficients}. This means that in such cases the measurements of the dust depletion are extrapolated from the general behaviour of the other environments, and our estimates on the dust composition are indicative. For the cases of targets that have less than three measurements of metals, the calculation of f$_{\rm{M}_{X}}$ is not reliable and thus we exclude them. Figures \ref{fig:dustcomp_all_qso_dla} to \ref{fig:dustcomp_all_smc} show the dust composition with respect to [Zn/Fe] for all the environments. The systems for which we have a measurement of dust depletion for each metal are shown with filled symbols and those calculated from the coefficients with empty symbols.



\subsection{Dust extinction $A_{V}$ from depletion}
\label{sec:dustext}
The extinction of light by dust grains depends on several factors, such as their density, their size distribution, and composition \citep[e.g.][]{Draine2003}. While typically the dust extinction $A_{V}$ is measured from the variations on the spectral continuum, it is also possible to estimate an $A_{V}$ from the dust depletion (and the dust-to-metals ratio), which we refer to as $A_{V, \rm depl}$. However, because dust extinction depends on the cross section of the dust grains, it is more sensitive to the number of grains, rather than the mass of the individual grains. Therefore, we derive the $A_{V, \rm depl}$ based on a DTM that is defined in terms of number of atoms (or column densities), which we refer to as DTM$_{\rm N}$.
We derive the DTM$_{\rm N}$ in terms of column density by number as,
\begin{equation}
\begin{aligned}
    \rm{DTM}_{\textit N} = \frac{\Sigma_{X_i}\, (1-10^{\delta_{X_i}})~10^{[X_i/H]_{\odot}}}{\Sigma_{X_i}\, 10^{[X_i/H]_{\odot}}},
\label{dtm_column}
\end{aligned}
\end{equation}
The DTM in terms of column density (DTM$_{\rm N}$) is on average 0.05 times lower than the DTM in terms of mass (DTM), as expected because Eq. \ref{dtm_column} is equivalent to Eq. \ref{dtm} without adding the atomic weights W$_{X}$ of the elements. However, the overall trend is the same. Figure \ref{fig:dtm-dtmN} shows the relation between the DTM by mass and the DTM by column density (DTM$_{\rm N}$). Our measurements of both DTM and DTM$_{\rm N}$ are given in Tables \ref{dlastable} - \ref{grbtable}}.

The dust extinction $A_{V}$ can be derived from the depletion estimated DTM$_{\rm N}$ as,

\begin{equation}
    A_{V,\rm depl} = \frac{\rm{DTM}_{\textit N}}{\rm DTM_{{\textit N,\rm Gal}}} \times \left(\frac{A_{V}}{\textit N(\rm H)}\right)_{Gal} \times \textit N(\rm H) \times 10^{[M/\rm{H}]_{tot}},
    \label{av}
\end{equation}
where $\left(\frac{A_{V}}{N(H)}\right)_{\rm Gal} = 0.45$ $\times$ 10$^{-21}$ mag~cm$^{2}$ is the Galactic DTM ratio from \citet{Watson2011}.  DTM$_{\rm{N,Gal}} = 0.354$ is the dust-to-metal ratio  for the Galaxy in terms of column density, derived using Eq. \ref{dtm_column} and assuming an average [Zn/Fe] = 1.22 for the Milky Way. N(H) is the total column density of H, defined as N(H) = N(\ion{H}{I}) + 2 N(\ion{H}{2}), although for most QSO- and GRB-DLAs only \ion{H}{I} is measured and \ion{H}{2} is negligible \citep[e.g., the mean molecular fraction measured in QSO-DLAs in][is log\,f$_{\ion{H}{2}} \sim $ -6.2]{Noterdaeme2008}. [\textit{M}/$\rm{H}$]$_{\rm{tot}}$ is the total dust-corrected metallicity. 

\begin{figure*}[ht!]
    \centering
    \includegraphics[width=\textwidth]{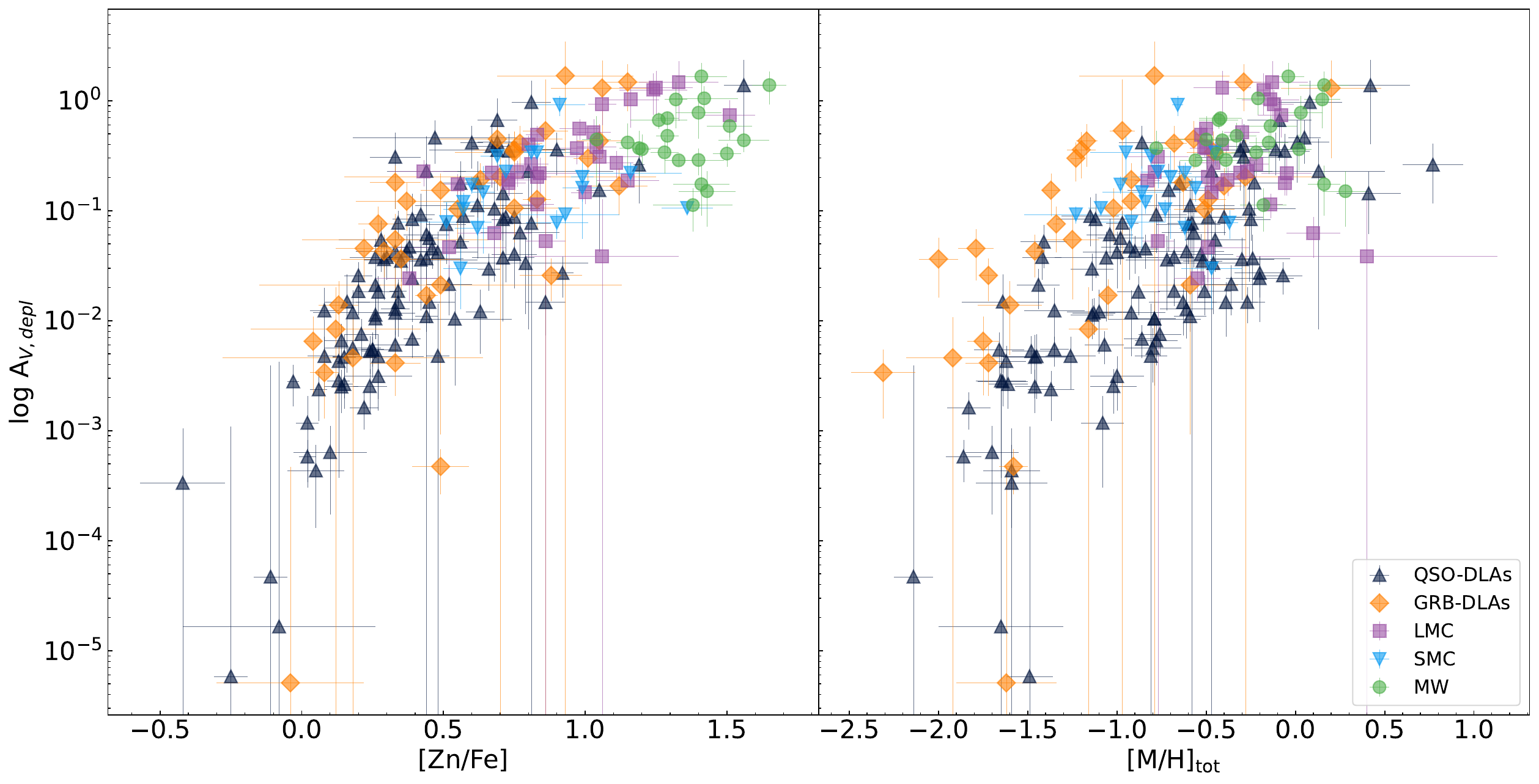}
    \caption{Dust extinction $A_{V,\rm depl}$ as a function of [Zn/Fe] (left panel) and as a function of the total dust-corrected metallicity [\textit{M}/$\rm{H}$]$_{\rm{tot}}$ (right panel). For the Milky Way metallicities $[M/\rm H]_{\rm{tot}}$ are from \citet{DeCia2021}.}
    \label{fig:av_znfe}
\end{figure*}

Figure \ref{fig:av_znfe} shows the distribution of $A_{V, \rm depl}$ with [Zn/Fe] (left panel) and with the total dust corrected metallicity [\textit{M}/$\rm{H}$]$_{\rm{tot}}$ (right panel). The DTM, DTG, DTM$_{\rm N}$ and $A_{V,\rm depl}$ results for all the environments are reported in Tables \ref{dlastable}- \ref{grbtable}.

\section{Discussion}
\label{sec: discussion}


\subsection{DTM, DTG, and the origin of dust}
\label{sec:dtmdtg2}

The evolution of the DTM and DTG ratios with metallicity gives us clues about the origin of interstellar dust. The three candidate mechanisms for the formation of the bulk of cosmic dust are dust condensation in the envelopes of AGB stars \citep{Gail2009}, condensation in SNe ejecta \citep{Dunne2003,Matsuura2015}, and grain growth in the ISM \citep{Draine2003,Mattsson2012,Dwek2016,DeCia2016}. 

Different dust production mechanisms can produce different types of dust. AGB stars produce C-rich dust as well as silicates, while SNe produce mostly silicates. Type Ia SNe are the major Fe producers but dust produced by Type Ia SNe has generally not been observed, with the exception of \citet{Dwek2016,Nagao2018}. ISM grain growth by condensation upon existing seeds produced by stars is the mechanism that can produce silicates but also the amounts of C- and Fe-dust observed. 

Different mechanisms of dust production have different dependencies on the environment metallicity. Dust that is mostly built up with ISM grain growth has DTM ratios that increase with metallicity, because the grain growth rate is metallicity dependent. On the other hand, SNe-produced dust should show a constant DTM \citep{Mattsson2014}, because dust production by SNe is weakly dependent on metallicity. AGB stars are often considered secondary for dust production, because of their longer lifetimes, although they cannot be neglected on short timescales \citep[$\sim$500~Myr,][]{Valiante2009}.



Although grain growth in the ISM can occur at any time during the lifetime of a galaxy, \citet{Zhukovska2008} show that there is likely a critical metallicity above which grain growth in the ISM becomes dominant over other sources of dust production. Their evolution model shows that at very low metallicities [\textit{M}/$\rm{H}$]$_{\rm{tot}}$ $\leq -2$ in the solar neighbourhood the grain growth timescale exceeds by about 10~Myr the average lifetime of the molecular cloud. This means that until this point, not a lot is added to the dust content of the interstellar matter. When the metallicity of the ISM has reached [\textit{M}/$\rm{H}$]$_{\rm{tot}} = -2$ some dust starts to condensate in the ISM and at [\textit{M}/$\rm{H}$]$_{\rm{tot}}$ $\sim -1$ the rate of growth increases rapidly and grain growth is the dominant mechanism of dust production in the diffuse ISM \citep{Zhukovska2008}. At high metallicities, such as in the Milky Way, the DTM ratio has a smaller dependence with metallicity. This is possibly because dust production depends not only on metallicity, but also temperature, density and pressure. In the Milky Way the presence of cold dense gas and high pressure provide favorable conditions for the formation of dust. In this case, due to the higher dependence of ISM grain growth on density and pressure, the metallicity dependence is lower \citep{Blitz2006}.

The top panel of Fig. \ref{fig:dtm_dtg_znfe} shows the DTM ratio with respect to the dust tracer [Zn/Fe] for the Milky Way, LMC, SMC, QSO- and GRB-DLAs. It is evident that there is a tight correlation between the DTM and [Zn/Fe], which is expected by the fact that the depletions that are used for the DTM calculation are linearly correlated with [Zn/Fe] (see Eq. \ref{deltaxcoeff}). The DTM is lower for QSO-DLAs and increases with increasing [Zn/Fe], i.e., for dustier systems, like the Milky Way.

The bottom panel of Fig. \ref{fig:dtm_dtg_znfe} shows the DTG with respect to [Zn/Fe] for the Milky Way, LMC, SMC, DLA- and GRB-DLAs. The DTG ratio is increasing with [Zn/Fe] and there is a larger scatter in the linear relation than what we see for the DTM (top panel of Fig. \ref{fig:dtm_dtg_znfe}). Fig. \ref{fig:dtm_dtg_mh} shows the DTM (top panel) and DTG (bottom panel) with respect to the total dust-corrected metallicity [\textit{M}/$\rm{H}$]$_{\rm{tot}}$. We find that overall the DTM is increasing with metallicity. For QSO-DLAs, we see a relatively sharp decline of the DTM at low metallicities (at  [\textit{M}/$\rm{H}$]$_{\rm{tot}} < -1$), indicating that the net dust production rate is lower at low metallicities. This decline of the DTM could indicate that dust is destroyed more efficiently than being formed in this low-metallicity regime. However, uncertainties are large for these few systems probably as a result of the low amounts of dust, which is difficult to measure with this method. 

The analytic dust production and evolution model tracks from \citet{Mattsson2020} are also shown in Fig. \ref{fig:dtm_dtg_mh}. The model is overplotted, using eq. 42 of \citet{Mattsson2020} and assuming that the dust destruction parameter $\delta$ is essentially equal to the growth parameter $\epsilon$. The model considers all sources of dust, i.e., stellar sources (AGB stars, SNe) and grain growth in the ISM, involves several simplifying assumptions and is based on an idealised model of grain growth in turbulent cold molecular clouds.
Some parameters are modified to fit the observational data well, that is, the stellar dust yield (y$_{d}$) is lowered by an order of magnitude with respect to \citet{Mattsson2020} and the solar abundance Z$_{\odot}$ is updated to the most recent value from \citet{Asplund2021}. The parameters for the model are then y$_{Z}$ = 0.01, y$_{d}$ = 1$\times$ 10$^{-4}$,  $\delta$ = $\epsilon$ and Z$_{\odot}$ = 0.0139. The shaded regions in the plots show the parts of the DTM-[\textit{M}/$\rm{H}$]$_{\rm{tot}}$ and DTG-[\textit{M}/$\rm{H}$]$_{\rm{tot}}$ planes that are covered by varying $\delta$ = $\epsilon$ from 1 to 16. Surprisingly, the predicted (stellar) dust yield is only 1$\%$ of the metals yield, while the previous estimates by \citet{Mattsson2020} indicated 10-20$\%$. This suggests that the overall stellar population (and not individual cases) produce much less dust than previously estimated. We note that with our data we probe the integrated metals and dust recycled by the stellar sources (AGB stars, SNe) in the warm diffuse ISM. We might be missing the dust produced by stars in the dense and dustier regions around AGB stars or SNe.

The fact that the DTM is increasing with metallicity indicates that grain growth in the ISM, is likely the dominant mechanism for dust production. This is supported by the fact that the data fit well the dust production and evolution model trend from \citet{Mattsson2020}. We note that due to the nature of the measurements used in this paper (column densities of the metals in the warm neutral gas from UV absorption-line spectroscopy), there is a selection effect that gives privilege to sampling of environments where there is dust growth in the (more diffuse) ISM. Moreover, with the depletion measurements we probably cannot distinguish between grain growth and dust destruction/sputtering in the diffuse ISM. With our method we do not directly probe the very dense envelopes of AGB stars, or regions of condensations in SNe ejecta, but any dust that has been produced or recycled into the warm neutral ISM. In addition the number of SNe or AGB stars along a line of sight depends on the star-formation history of the local environment sampled. As shown in Paper I, the depletions sequences are independent of the star-formation histories of the different galaxies that were sampled. This means that the total budget of dust grains in the warm neutral medium does not depend on the star-formation histories of the galaxies. However, our technique does not probe dust in the densest regions of the ISM (cold and molecular neutral medium) and in particular C-rich dust grains, which contribute significantly to the dust budget.

We find that the increase of the DTM with metallicity is not developing with the same rate for all the environments (49\,$\%$ increase for QSO-DLAs, 40\,$\%$ for GRB-DLAs, 23\,$\%$ for the LMC, 21\,$\%$ for the SMC and 23\,$\%$ for the Milky Way). In environments that have high metallicity and large amounts of cold dense gas (high pressure), like in the Milky Way disk, the dependence on the metallicity seems to become less important. In this case, grain growth in the ISM depends more strongly on pressure and thus molecular production and ISM grain growth \citep{Blitz2006}. This can cause a shallower increase in the DTM with metallicity for environments with higher pressure.

Overall, the DTM is larger for dustier and higher metallicity systems, like the Milky Way and lower for QSO-DLAs. An increase of the DTM with metallicity has been observed in local and distant galactic environments \citep[e.g.][]{DeCia2013,DeCia2016,Wiseman2017}, with this tendency somehow flattening above 0.1\,Z$_{\odot}$ and reaching the Milky Way values. However, there are differences between the DTM calculations in these works. In our method, we use all the metals that contribute into dust, including C, by using the depletion of C that is estimated either from measurements for the Milky Way or from its empirical relation with [Zn/Fe] (calibrated with Milky Way data only) for the rest of the environments. We also calculate the DTM in terms of mass, by weighting each element for its atomic weight. \citet{DeCia2013} and \citet{DeCia2016} derive DTM$_{\rm N}$ in terms of column density and only based on the depletion of Fe. \citet{Wiseman2017} calculate the DTM$_{\rm N}$ in terms of column density and not including C.All these works refer to a normalized DTM$_{\rm N}$ with respect to the Milky Way. If we consider this, our DTM$_{\rm N}$ is consistent with the DTM$_{\rm N}$ estimated by \citet{Wiseman2017} within the uncertainties. 

In Fig. \ref{fig:dtm_dtg_mh}, the DTM suggests that GRB-DLAs are dustier than QSO-DLAs for equal amounts of metals, while in Figs. \ref{fig:dtm_dtg_znfe} and \ref{fig:dtm_dtg_mh} the DTG indicates that GRB-DLAs exhibit more gas with respect to QSO-DLAs along the lines of sight.
In addition, most systems lying above the model expectations in Fig \ref{fig:dtm_dtg_mh} are GRB-DLAs at low metallicity, while those below are QSO-DLAs at [\textit{M}/$\rm{H}$]$_{\rm{tot}} > -0.3$. A possible explanation for these effects is the increased amount of colder dense gas and pressure in GRB hosts with respect to the more diffuse and warmer gas probed by QSO DLAs. This is vastly a geometrical effect of GRB lines of sight crossing through more central parts of their host galaxy, while QSO-DLAs probe more peripheral regions of the absorbing galaxy \citep[e.g.,][Krogager et al. in prep]{Fynbo2008}. Overall, our results suggest that the production of dust is not only dependent on the metallicity - or the availability of metals - but also on gas temperature, pressure and density in each environment. The scatter that we observe in the DTM values at a given metallicity may be due to the different dynamical timescales that the stellar and ISM sources contribute to dust at various redshifts. At higher redshifts but for the same [\textit{M}/$\rm{H}$]$_{\rm{tot}}$ it is possible that some sources did not have enough time to form dust in such short timescales (the Universe is just $\sim$ 1\,Gyr old at $z$ = 6). The  scatter in our DTM values suggest a complex interplay of factors within galactic environments. With our method we primarily probe the characteristics of dust within the diffuse ISM. We acknowledge that our approach may not fully capture contributions from specific stellar sources, such as carbonaceous dust from AGB stars, in denser regions. It is indeed possible that the scatter observed in our DTM values arises from the combined effects of sampling different combinations of more diffuse and denser ISM, as well as regions with varying star-formation histories. The results presented in Paper I indicate that in the diffuse ISM, dust origin appears to be less dependent on the star formation history of galaxies. As our study predominantly focuses on the diffuse ISM, we recognize the need for further research to explore methods that better represent denser regions, which may be influenced by variations in star formation history. The DTM overall tends to decrease with redshift, which is likely due to the dependence of the DTM on the decreasing metallicity \citep{Heintz2023}.

The DTG shows a clear increasing trend with metallicity for all the environments (see bottom panel of Fig. \ref{fig:dtm_dtg_mh}). The increase of DTG with metallicity has been consistently observed in previous works \citep{Issa1990,Lisenfeld1998, Draine2007, Galliano2008,Remy-Ruyer2014, DeVis2017, Galliano2018,Li2019,Roman-Duval2022iv}. The observed DTG trend with metallicity is in agreement with the dust production and evolution model from \citet{Mattsson2020}. \citet{Roman-Duval2022iv} compare their observed DTG-metallicity trend with the model tracks from \citet{Feldmann2015}, which fits well the FIR observations of galaxies from \citet{Remy-Ruyer2014}. In this comparison, these author find a discrepancy of the QSO-DLA observations with the FIR observations from \citet{Remy-Ruyer2014}, which is still present in our current observational dataset. This discrepancy highlights that FIR observations probe a very different kind of environment, likely the most dense clouds around star-forming regions and AGB stars that may have dust with a different nature from what is observed in the diffuse ISM. The DTG-metallicity trend that we observe is consistent with the one observed in \citet{Roman-Duval2022iv}.


\subsection{Cosmic metallicity evolution of the neutral gas - Comparison between QSO- and GRB-DLAs}
\label{sec:cosmic_metal_evolution}

\begin{figure*}[!ht]
    \centering
    \includegraphics[width=\textwidth]{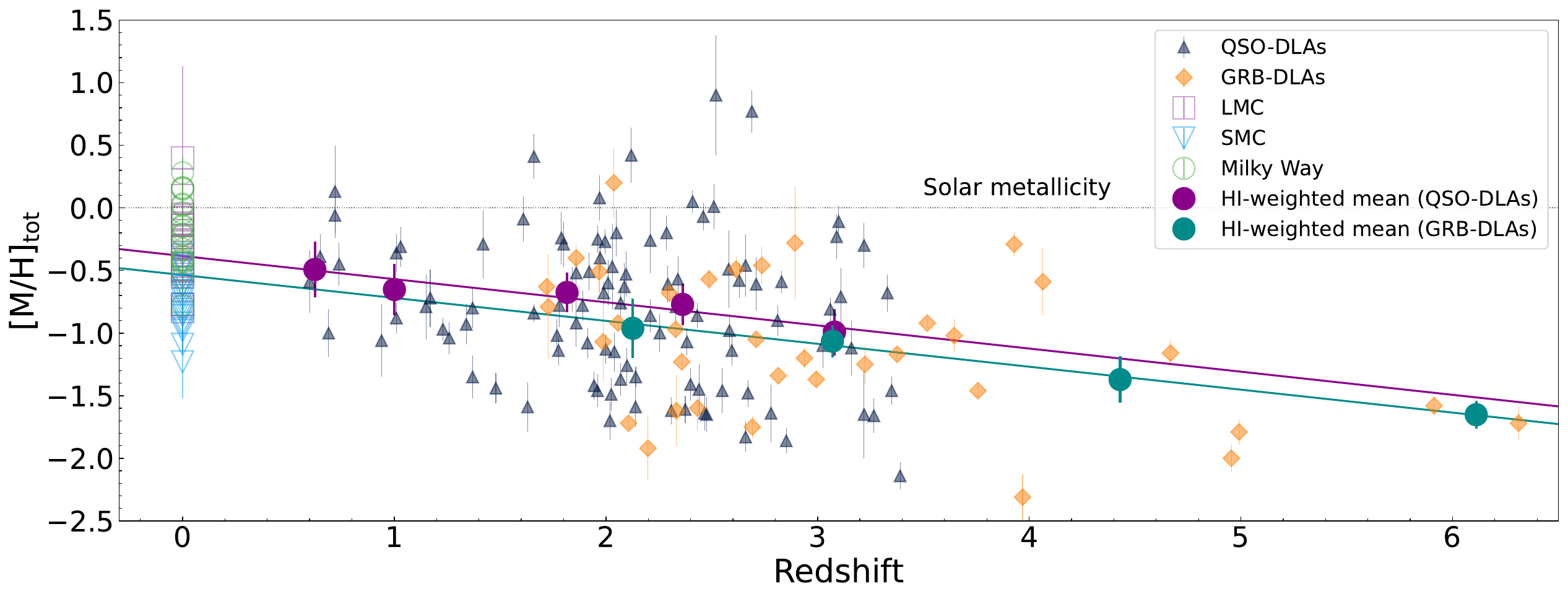}
    \caption{Total dust-corrected metallicity [\textit{M}/$\rm{H}$]$_{\rm{tot}}$ evolution with redshift $z$ for GRB-DLAs (orange diamonds) and QSO-DLAs (black triangles). The larger points (green for GRB-DLAs and purple for QSO-DLAs) show the HI-weighted mean metallicities binned in redshift and the lines are their corresponding linear fits. The open symbols at redshift $z = 0$ refer to the Milky Way (green), the LMC (purple) and the SMC (blue). The dotted line shows the solar metallicity ([\textit{M}/$\rm{H}$]$_{\rm{tot}} =0$).}
    \label{fig:weighted_mean}
\end{figure*}

In Fig. \ref{fig:weighted_mean} we compare the metallicity evolution with redshift between QSO- and GRB-DLAs. We calculate the $\ion{H}{i}$-weighted mean from the linear metallicities in five redshift bins for QSO- and GRB-DLAs and plot the points at the mean redshift. We fit the $\ion{H}{i}$-weighted mean metallicities, taking into account the y-axis uncertainties. The individual metallicity measurements in different environments are shown in Fig. \ref{fig:weighted_mean}. The two linear fits result to [\textit{M}/$\rm{H}$]$_{\rm{tot}}$ $= (-0.38 \pm 0.07) + (-0.18 \pm 0.03)\times z$ for QSO-DLAs and [\textit{M}/$\rm{H}$]$_{\rm{tot}}$ $= (-0.53 \pm 0.05) + (-0.18 \pm 0.01)\times z$ for GRB-DLAs, where $z$ is the redshift. There is a small difference between the fits, which we quantify by calculating the difference between the reduced $\chi_{\nu}^2$ of the two fits ($\Delta \chi_{\nu}^2 = 0.04$). We note that the two samples have different redshift ranges, with most of the targets concentrating in the redshift range $1.8 < z < 3.2$. In this range, the observed scatter is large, and is reflecting the diversity in metallicity of galaxies at any given time. This is not complete, however, since optically- or UV-selected QSOs tend to have lower [\textit{M}/$\rm{H}$]$_{\rm{tot}}$, because their selection is biased by the presence of dust \citep{Krogager2019}. At higher and lower redshifts there are less data points due to observational limitations. The effect of dust bias will be further investigated with the 4MOST–Gaia Purely Astrometric Quasar Survey (4G-PAQS) \citep{Krogager2023}.


The $\ion{H}{i}$-weighted mean metallicity is a measure of the metallicity in the neutral gas in the Universe, rather than the metallicity of the individual systems. The agreement between the two fits suggests that QSO- and GRB-DLA may be arising from a similar population of galaxies, or at least probe the neutral gas in the Universe in a similar way (see also Krogager et al. in prep.). We find a decrease of neutral gas metallicity with redshift ($0.6 < z < 6.3$), in agreement with previous studies \citep[e.g.][]{DeCia2018,Mattsson2019, Peroux2020}. The inclusion of GRB-DLAs extends the study of the cosmic chemical evolution to $z>6$ \citep[]{Heintz2023, Saccardi2023}.

\subsection{Dust composition}
\label{sec:dustcomp2}

Figure \ref{fig:dust_composition} shows the evolution of dust composition derived from the dust depletion, with [Zn/Fe] for the elements that contribute at least 1\,$\%$ to the dust composition (f$_{\rm{M}_{X}}$> 1$\%$). Dust depletion can be used to get information on the dust composition in various ways \citep{Savage1996,Jenkins2014,DeCia2016,Mattsson2019,Roman-Duval2022a}.

\begin{figure*}[h!]
\centering
\includegraphics[width=0.75\textwidth]{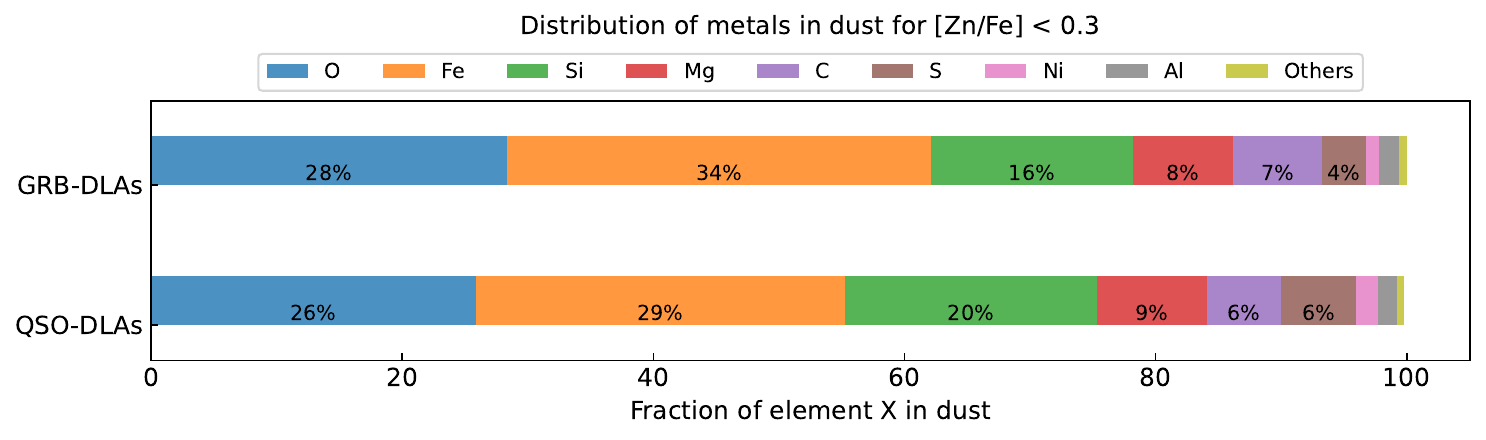}
\includegraphics[width=0.75\textwidth]{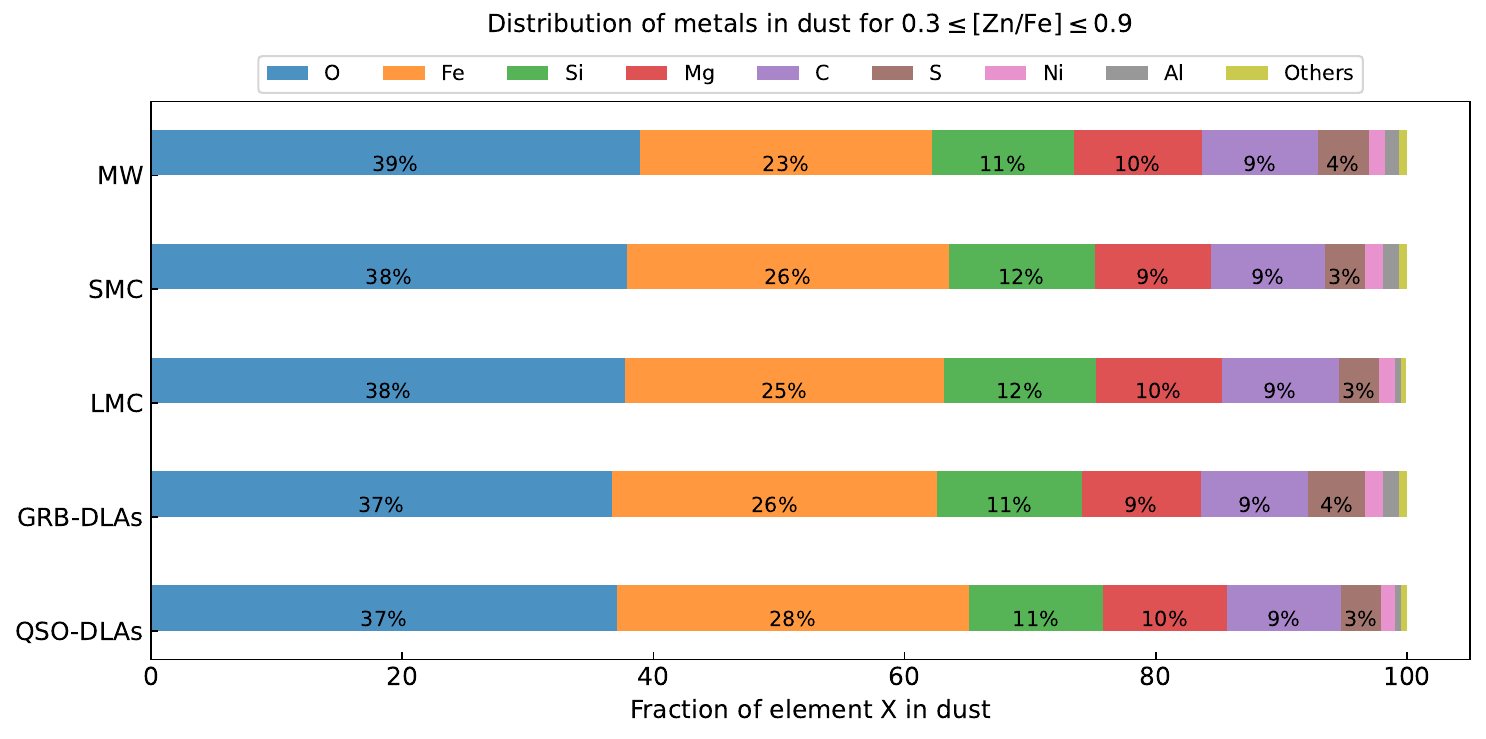}
\includegraphics[width=0.75\textwidth]{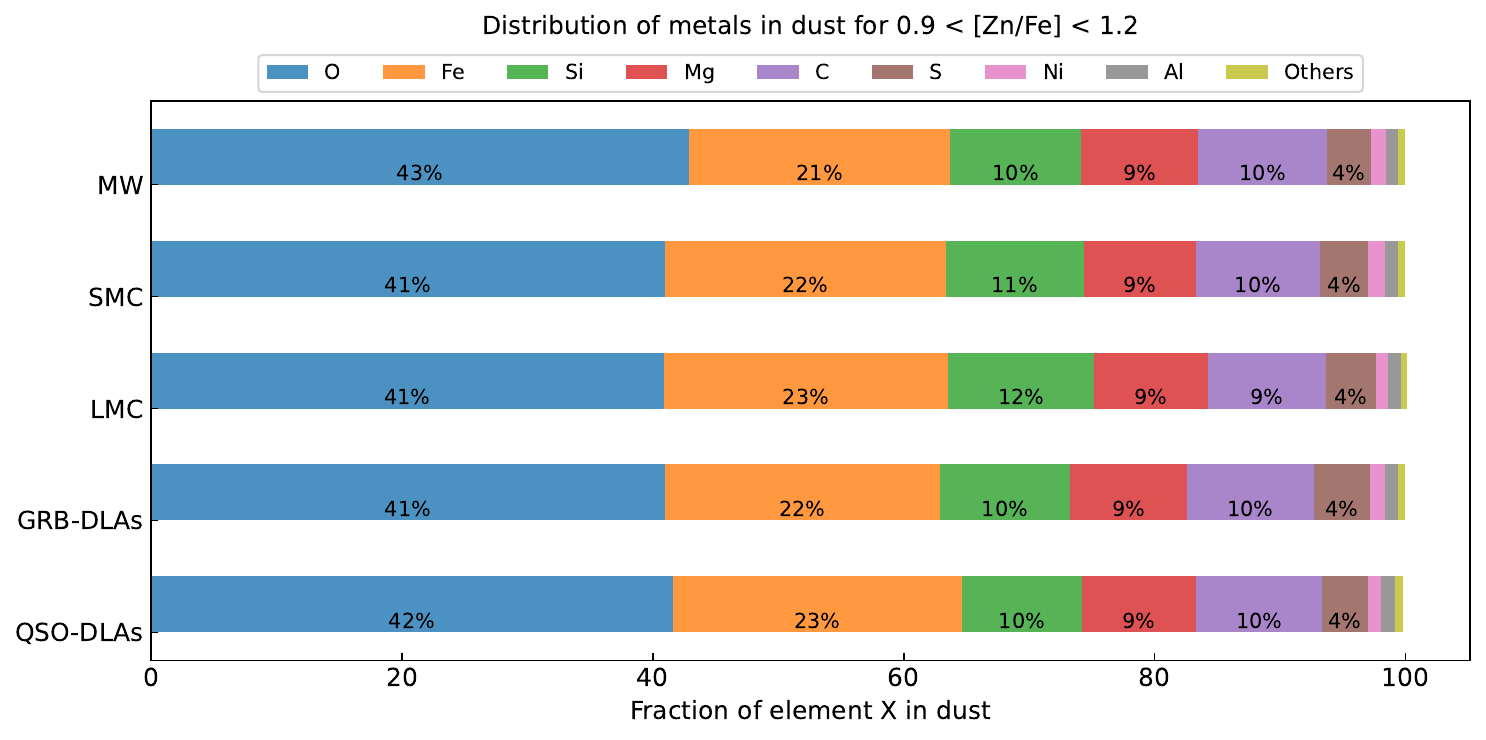}
\caption{Distribution of metals that compose dust for the main elements (f$_{\rm{M}_{X}}$ > 1$\%$) that contribute into dust. The top panel shows the distribution of elements for [Zn/Fe] < 0.3 in QSO- and GRB-DLAs, the middle panel for 0.3 < [Zn/Fe] < 0.9 in all the environments and the bottom panel for 0.9 < [Zn/Fe] < 1.2.}
\label{fig:donuts}
\end{figure*}

In each panel of Fig. \ref{fig:dust_composition} we can see that the main components of dust are by mass, O, Fe, Si, Mg, C, S, Ni and Al. The distribution of the elements that compose dust varies, however, with the amount of dust in the system measured with [Zn/Fe]. Figure \ref{fig:donuts} visualizes this distribution of metals that compose dust for the least dusty systems (upper panel, e.g. [Zn/Fe] < 0.3), the distribution of metals in the middle range of [Zn/Fe] (middle panel, e.g. 0.3 $\leq$ [Zn/Fe] $\leq$ 0.9) and for most dusty systems (lower panel, e.g. 0.9 < [Zn/Fe] < 1.2). Systems with low dust content ([Zn/Fe] < 0.3) are observed among QSO- and GRB-DLAs, which tend to have a low metallicity. In these systems, the most abundant elements in dust are Fe, O, and after these Si. On the contrary, in the average [Zn/Fe] and high dust content regime (0.9 < [Zn/Fe] < 1.2) O is more abundant than Fe in all the galactic environments. Overall, in GRB-DLAs and QSO-DLAs the contribution of Fe and Si to dust increases steeply as we go towards lower [Zn/Fe] (i.e. less dust), while O decreases. We speculate that this least dusty regime may be one where the seeds of dust grains did not undergo as much grain growth by the gradual accumulation of matter in the ISM, because there was not enough time, or underwent a more efficient dust destruction process by SNe shocks. In low metallicity and low dust environments the dust grains are small and are more easily destroyed by SNe shocks.

The main dust destruction processes in the ISM are either from sputtering by high-velocity SN shocks and photo-destruction by high-energy photons \citep{Dwek1996,Jones1996,Bocchio2014,Slavin2015}, or from consumption of interstellar dust by star formation (astration).

In Table \ref{ppm_tab} we present the gas- and solid phase abundances in parts-per-million (ppm) of the main elements that contribute to the dust composition for systems in different environments and for the three different intervals of [Zn/Fe] defined above. At low-depletion levels ([Zn/Fe] < 0.3), where dust starts to form, we find for QSO- and GRB-DLAs that the O abundance in dust is $\sim$ 3 times higher than the Si abundance. This suggests that the presence of pyroxene (\ch{Mg_{x}Fe_{(1-x)}SiO3}) is likely. The amounts of Fe and Mg observed in the least-dusty regime suggest a significant contribution of enstatites (\ch{MgSiO3}) and metallic Fe. There is no need to invoke oxydation of Fe (e.g. \ch{FeO}) in the least dusty regime, because all the oxygen can in principle be included in pyroxenes. A more detailed analysis of the dust composition, such as the one presented in \citet{Mattsson2019}, is required to accurately constrain the dust composition of the different environments.

\begin{table*}[!h]
\caption{Gas- and solid-phase abundances of the main elements that contribute to the dust composition for systems in different environments and for different intervals of  [$\rm{Zn/Fe}$].}
\label{ppm_tab}
\centering
\tiny
\begin{tabular}{c|c|ccccc|c} 
 \hline
 \multirow{3}{*}{Element X} &
 \multirow{3}{*}{[\rm X/H]$_{\rm gas}$ (ppm)} 
 & \multicolumn{5}{c|}{[\rm X/H]$_{\rm dust}$ (ppm)} &  \multirow{3}{*}{Interval [$\rm{Zn/Fe}$]}\\
 & &&&&&&\\
& & QSO-DLAs & GRB-DLAs & Milky Way & LMC & SMC \\  
\hline\hline
O  & 490 & 35 & 27 & ... & ... & ... & [$\rm{Zn/Fe}$] < 0.3 \\
& & 102 & 118 & 154 & 126 & 125 & 0.3 $\leq$ [$\rm{Zn/Fe}$] $\leq$ 0.9 \\
& & 186 & 178 & 181 & 172 & 168 & 0.9 < [$\rm{Zn/Fe}$] < 1.2 \\
\hline
Fe & 29 & 10 & 27 &  ... & ... & ... & [$\rm{Zn/Fe}$] < 0.3 \\
& & 23 & 23 & 26 & 25 & 25 & 0.3 $\leq$ [$\rm{Zn/Fe}$] $\leq$ 0.9\\
& & 27 & 27 & 27 & 27 & 27 & 0.9 < [$\rm{Zn/Fe}$] < 1.2 \\
\hline
Si & 32 & 13 & 8 &  ... & ...&... & [$\rm{Zn/Fe}$] < 0.3 \\
& & 14 & 20 & 25 & 25 & 22 & 0.3 $\leq$ [$\rm{Zn/Fe}$] $\leq$ 0.9 \\
& & 21 & 25 & 27 & 26 & 27 & 0.9 < [$\rm{Zn/Fe}$] < 1.2 \\
\hline
Mg & 35 & 7 & 5 &  ... & ... & ... & [$\rm{Zn/Fe}$] < 0.3 \\
& & 18 & 20 & 27 & 24 & 20 & 0.3 $\leq$ [$\rm{Zn/Fe}$] $\leq$ 0.9 \\
& & 24 & 27 & 28 & 22 & 26 & 0.9 < [$\rm{Zn/Fe}$] < 1.2 \\
\hline
C & 288 & 10 & 8 &  ... & ... & ... & [$\rm{Zn/Fe}$] < 0.3 \\
& & 34 & 37 & 48 & 43 & 40 & 0.3 $\leq$ [$\rm{Zn/Fe}$] $\leq$ 0.9\\
& & 56 & 33 & 81 & 48 & 54 & 0.9 < [$\rm{Zn/Fe}$] < 1.2 \\
\hline
S & 14 & 3 & 2 &  ... & ... & ... & [$\rm{Zn/Fe}$] < 0.3 \\
& & 5 & 6 & 8 & 6 & 5 & 0.3 $\leq$ [$\rm{Zn/Fe}$] $\leq$ 0.9\\
& & 7 & 10 & 8 & 8 & 9 & 0.9 < [$\rm{Zn/Fe}$] < 1.2 \\
\hline
Ni & 2 & 1 & 0.3 &  ... & ... & ... & [$\rm{Zn/Fe}$] < 0.3 \\
& & 1 & 1 & 1 & 1 & 1 & 0.3 $\leq$ [$\rm{Zn/Fe}$] $\leq$ 0.9 \\
& & 1 & 1 & 2 & 1 & 1 & 0.9 < [$\rm{Zn/Fe}$] < 1.2 \\
\hline
Al & 3 & 1 & 1 &  ... & ... & ... & [$\rm{Zn/Fe}$] < 0.3 \\
& & 2 & 2 & 3 & 3 & 2 & 0.3 $\leq$ [$\rm{Zn/Fe}$] $\leq$ 0.9 \\
& & 3 & 3 & 3 & 2 & 3 & 0.9 < [$\rm{Zn/Fe}$] < 1.2\\
\hline
\end{tabular}
\end{table*}



The different environments considered in this study span different ranges of dust content ([Zn/Fe], see Fig. \ref{fig:dust_composition}). Here we discuss their average properties. For the full range of [Zn/Fe], for all the environments, we find that the largest fraction of dust mass is contributed by O and is increasing with [Zn/Fe] with a smaller increase in the Milky Way and the largest increase for QSO-DLAs. The largest amount of O in dust is found in the Milky Way (44\,$\%$) and the smallest in GRB-DLAs (35\,$\%$). The second most abundant element in dust is Fe, which is decreasing with [Zn/Fe] for all the environments. A possible cause for the opposite evolutions of the mass abundance of O and C with respect to Fe and other refractory elements is the following. For the least dusty systems, at low levels of [Zn/Fe], most of the Fe and other refractory elements are available in the gas-phase and they are the ones that can more easily be incorporated in dust grains. On the other hand, for very dusty systems, at high levels of dust depletion, there are much fewer atoms of Fe and refractory metals available in the gas-phase to form additional dust, while O and C are always very abundant in the gas-phase. Layers of complexity will need to be added to this simplified picture when considering systems that have intrinsically very low abundance of Fe and heavily refractory metals, for example in the very distant Universe.  The chemical evolution of galactic environments and their local star-formation histories influence the availability of metals for dust formation. In regions with lower metallicity and lower dust content, which are chemically less evolved, there might be a reduced abundance of heavy elements crucial for dust production.
Si, C, Mg, S, Ni and Al have smaller contributions in dust (< 12$\%$) in all the environments.

Our current data suggest that C has a similar mass abundance in dust for all the environments, 8\,$\%$, 9\,$\%$, 10\,$\%$, 10\,$\%$, 12\,$\%$ for QSO-DLAs, GRB-DLAs, LMC, SMC and Milky Way respectively. However, the Milky Way and Magellanic clouds have substantially different extinction curves, in particular the presence and strength of the 2175$\,\AA$ feature \citep[e.g.][]{Pei1992}. The origin of the 2175$\,\AA$ feature remains unclear and it is likely that a combination of several species that produce it. Some candidates that may produce the feature have been proposed, such as non-graphitic C \citep{Mathis1994} or polycyclic aromatic hydrocarbons \citep[PAHs,][]{Draine2003}. This indicates that the Milky Way has a much higher content of C-rich dust, possibly polycyclic aromatic hydrocarbons (PAH) \citep{Shivaei2022}, than the LMC and even more the SMC. The SMC extinction curve does mostly not show the 2175$\,\AA$ feature, with few exceptions \cite[e.g.][]{Gordon1998,Gordon2003,Welty2010}. Most GRB afterglows also do not show a 2175$\,\AA$ feature in their extinction curve \citep[e.g.][]{Zafar2011,Schady2012}, with some exceptions \citep{Eliasdottir2009, Zafar2012, Heintz2019b}. Thus, one could expect a smaller contribution of PAH grains, and possibly C-rich dust and smaller grain sizes to their dust composition. The distribution of PAHs may actually vary with location, being more strongly associated with molecular clouds \citep{Sandstrom2010}. The dust depletion analysis probes more efficiently the warm neutral medium in galaxies, and may be not very sensitive to dust located in dense molecular clouds. Thus our results on the dust composition are only representative of the dust in the warm neutral medium, while the dust located in dense molecular clouds might have a different composition.

The depletion of C in dust is very uncertain for GRB- and QSO-DLAs, because the typical $\ion{C}{ii}$ line used to measure the C content is normally highly saturated. In Paper I we derived the depletion of C based only on five Milky Way measurements. This can only give a rough estimate of the C dust depletion. We extrapolate these estimates on the C depletion for the other environments and this is probably the cause of the similar percentages of C dust composition that we obtain, regardless of the presence or not of strong C-dust features in their extinction curves (i.e. SMC, LMC). This highlights the importance of a detailed study of the C depletion and the need for more and reliable measurements for systems in all environments.

Similar C fractions in dust can also result from the fact that most of C is produced by C-rich AGB stars \citep{Gustafsson1999,Mattsson2010}, the presence of which does not depend on the environment. Although, C-rich massive stars may significantly contribute to the C budget \citep{Gustafsson2022}.



\begin{table}[!h]
\caption{Average percentage of the main elements that contribute to the dust composition for systems in different environments.}
\centering
\small
\begin{tabular}{cccccc}
\hline\hline 
X & QSO-DLAs & GRB-DLAs & LMC & SMC & Milky Way\\
\hline
O & 36\,$\%$  & 35\,$\%$ & 38\,$\%$ & 39\,$\%$ & 44\,$\%$\\
Fe & 27\,$\%$  & 27\,$\%$ & 23\,$\%$ & 24\,$\%$ & 19\,$\%$\\
Si & 11\,$\%$  & 12\,$\%$ & 12\,$\%$ & 11\,$\%$ & 10\,$\%$\\
Mg & 9\,$\%$  & 9\,$\%$ & 10\,$\%$ & 9\,$\%$ & 9\,$\%$\\
C & 8\,$\%$  & 9\,$\%$ & 10\,$\%$ & 10\,$\%$ & 12\,$\%$\\
S & 4\,$\%$  & 4\,$\%$ & 4\,$\%$ & 3\,$\%$ & 4\,$\%$\\
Ni & 2\,$\%$  & 1\,$\%$ & 1\,$\%$ & 1\,$\%$ & 1\,$\%$\\
Al & 1\,$\%$  & 1\,$\%$ & 1\,$\%$ & 1\,$\%$ & 1\,$\%$\\
\hline
\label{percentdustcomp}
\end{tabular}
\end{table}


\subsection{Dust extinction from depletion compared to extinction from SED}
\label{sec:dustex}

Figure \ref{fig:av_znfe} shows the distribution of $A_{V,\rm depl}$ with [Zn/Fe] (left panel) and with the total dust corrected metallicity [\textit{M}/$\rm{H}$]$_{\rm{tot}}$ (right panel).
The $A_{V,\rm depl}$ for QSO-DLAs is generally small and is increasing with the dust tracer [Zn/Fe] and with the total dust corrected metallicity [\textit{M}/$\rm{H}$]$_{\rm{tot}}$. At higher metallicities and [Zn/Fe], the dust extinction is also expected to be higher. The right panel of Fig. \ref{fig:av_znfe} shows that, for a given metallicity, QSO-DLAs tend to have a lower $A_{V,\rm depl}$ than the other systems. This is probably due to the fact that QSO-DLAs probe outer galactic regions than GRB-DLAs \citep{Prochaska2007,Fynbo2008} and absorbing systems towards OB stars in local galaxies.

\begin{figure*}[h!]
\centering
\includegraphics[width=0.8\textwidth]{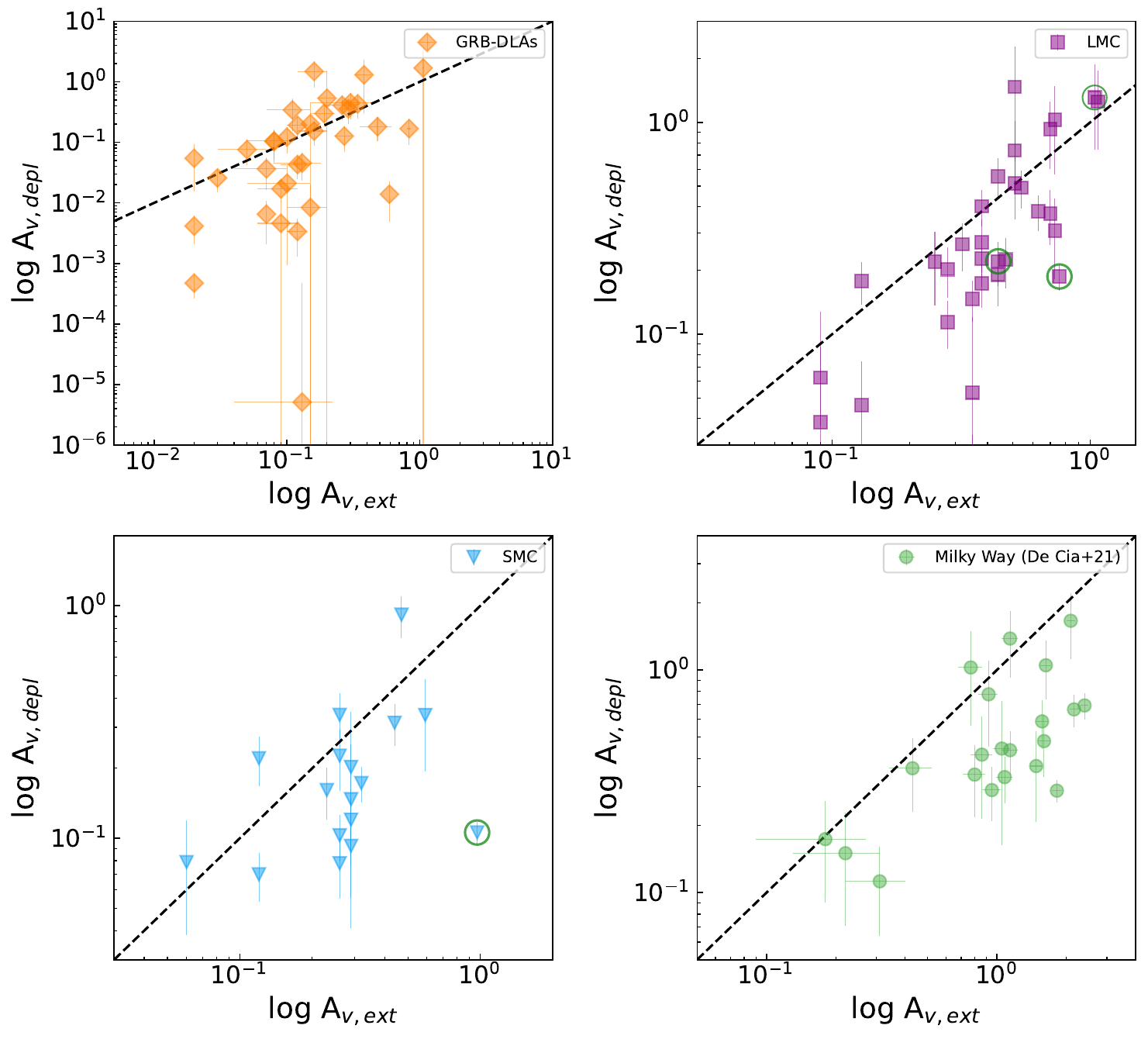}
\caption{Dust extinction $A_{V}$ as measured from extinction against the $A_{V}$ calculated using the dust depletion for GRB-DLAs, LMC, SMC and the Milky Way. The black dashed line indicates the 1:1 conversion between the two. The green circles show targets with a 2175$\,\AA$ feature or signs of carbonaceous dust.}
\label{fig:av_all}
\end{figure*}

We compare the $A_{V,\rm depl}$ with $A_{V,\rm ext}$, which we measured from the reddening $A_{\rm V} = E(B-V)$ $\times$ $R_{V}$, assuming $R_{V}$ = 3.08, for the Milky Way , $R_{V}$ = 3.16 for the LMC and $R_{\rm V} = 2.93$ for the SMC \citep{Pei1992}. In the case of GRB-DLAs we compare the $A_{V,\rm depl}$ with the $A_{V,\rm ext}$ measured from the SED in \citet{Heintz2023}. The comparison between $A_{V,\rm depl}$ and $A_{V,\rm ext}$ is shown in Fig. \ref{fig:av_all}. The green circles in \ref{fig:av_all} highlight systems in the Magellanic Clouds that show a 2175\,$\AA$ feature in their dust extinction curve. $A_{V,\rm ext}$ for QSO-DLAs are difficult to obtain and no measurements are available for the systems that we are studying here. The dust extinction $A_{V, \rm ext}$ of a sample of QSO-DLAs has been measured in previous studies \citep{Krogager2016,Heintz2016}, but is only possible for strong absorbers and/or for cases with the 2175\,$\AA$ bump present in their spectra. We find that $A_{V,\rm depl}$ is much lower than $A_{V,\rm ext}$ for the Milky Way. This could be due to the fact that with $A_{V,\rm depl}$ we are probing only the warm diffuse ISM, while $A_{V,\rm ext}$ is the cummulative extinction along the full line of sight, including the warm ISM and also regions of cold dense gas. On the other hand, for the LMC, SMC and GRB-DLAs the $A_{V,\rm depl}$ is generally in good agreement with $A_{V,\rm ext}$, with some exceptions that lie above or below the 1:1 line of equality. In the LMC and the SMC most of the systems lie below the line of equality ($A_{V,\rm ext}$ > $A_{V,\rm depl}$). We quantify the correlation between the two measurements of $A_{V}$ by calculating the $\chi_{\nu}^2$ between $A_{V,\rm ext}$ and $A_{V,\rm depl}$ for all the environments. The Milky Way has the highest $\chi_{\nu}^2$ with $\chi_{\nu}^2$ = 2.18 and then the SMC with $\chi_{\nu}^2$ = 2.14. GRB-DLAs have $\chi_{\nu}^2$ = 1.03 and finally the LMC $\chi_{\nu}^2$ = 0.63. We additionally estimate the Pearson's correlation coefficients r and report: r = 0.55 (GRBs), r = 0.71 (LMC), r = 0.30 (SMC), r = 0.48 (Milky Way). These indicate that the LMC points have the tightest correlation and the SMC the smallest. However, the low r for the SMC is driven only by the outlier marked in the green circle, which if excluded would give r = 0.61. Because this point has different properties due to the presence of the 2175$\,\AA$ bump in the dust extinction curve for this system and is a physical outlier.

\citet{Savaglio2004,Wiseman2017,Bolmer2019} also compare their estimates of the dust extinction derived from depletion $A_{V,\rm depl}$ with the dust extinction derived from SED ($A_{V,\rm ext}$) for GRB-DLAs. They find that some systems lie above or below the line of equality, with an overall larger scatter that is observed in this paper, and they report that most lines of sight have a larger $A_{V,\rm depl}$ than $A_{V,\rm ext}$. 
The difference in our method is that we include C, as well as several more metals in the derivation of the $A_{V,\rm depl}$ (all those that have an elemental abundance $12 + \rm log\,(X/H) > 3$). Dust extinction depends on the composition of the dust,
but also on the grain size distribution \citep{Mathis1977,Pei1992,Hoffman2016,Mattsson2020}, and gray dust has previously been suggested \citep[e.g.,][]{Wiseman2017} as a solution for the high  $A_{V,\rm depl}$, but low reddening in some sources. If grains are big enough, then the extinction depends less on the wavelength and more on the geometrical cross section, which has been observed in some cases \citep[e.g.,][]{Friis2015}. In this case, the extinction curve is achromatic and $A_{V,\rm ext}$ would be lower because each measurement is sensitive only to wavelength dependent extinction. On the other hand, $A_{V,\rm depl}$ is an average of the dust extinction in the warm neutral medium of the absorbing galaxies and would be more sensitive to gray dust. The presence of gray dust would result to $A_{V,\rm depl}$ > $A_{V,\rm ext}$. Most of the $A_{V,\rm ext}$ measurements for GRB-DLAs are done adopting an X-ray spectral slope \citep[e.g.,][]{Heintz2019b} and this makes them more robust in presence of gray dust. In any case, we observe only a few systems with significant $A_{V,\rm depl}$ > $A_{V,\rm ext}$, which possibly limits the effects of gray dust in this work.

The shape of the extinction curves depend on several properties, and can be steeper at lower metallicities \citep[e.g.][]{Shivaei2020}, like for GRB-DLAs. $A_{V,\rm ext}$ and $A_{V,\rm depl}$ may be probing different types of dust and different regions with lower or higher densities.
The methods used to measure $A_{V,\rm ext}$ and $A_{V,\rm depl}$ are very different, so that $A_{V,\rm ext}$ represents a sum of the extinction integrated along the line of sight, including clumps of dense cold regions and intervening systems at lower $z$, while $A_{V,\rm depl}$ represents an average value in the warm neutral medium through galaxies. 
On the other hand, the composition of dust plays a role in dust extinction, because the $A_{V,\rm ext}$ is very sensitive to C-rich dust, but the depletion of C is challenging to constrain. For the calculation of $A_{V,\rm depl}$ we do take into account C. However, only five Milky Way measurements are available (see Fig. B.2 of Paper I). These are based on only one weak transition ($\lambda\,$2325$\,\AA$), since the allowed transitions at $\lambda\,$1036$\,\AA$ and $\lambda\,$1335$\,\AA$ are strongly saturated \citep{Jenkins2009}. In Paper I we adopted the laboratory oscillator strength measurements with no additional correction for the oscillator strength of C.

In Paper I using the depletion coefficients we extrapolated the C depletion measured from a few Milky Way systems to the other environments where no measurements are available. This could mean that the depletion of C might be underestimated and might lead to a lower $A_{V,\rm depl}$. On the other hand, $A_{V,\rm ext}$ is estimated probing also C grains, especially in the Milky Way that has the strongest 2175$\,\AA$ feature. This could partly explain the higher values of $A_{V,\rm ext}$ and the large discrepancy with $A_{V,\rm depl}$ for the cases where C is present, like for most of the targets in the Milky Way. 

In fact, the presence of C-rich dust confined in molecular clouds may explain the discrepancy between $A_{V,\rm depl}$ and $A_{V,\rm ext}$ in a few outliers in the Magellanic Clouds as well as in the Milky Way (see Fig. \ref{fig:av_all}). The highly deviating points, Sk\,143 (SMC) at $A_{V,\rm ext}$ = 0.97 and Sk\,672 (LMC) at $A_{V,\rm ext}$ = 0.76, that are marked with green circles in Fig. \ref{fig:av_all}, are moderately reddened stars and the lines of sight with the highest molecular fraction and CN absorption within the SMC and LMC sample \citep{Cartledge2006,Welty2006}. Sk\,143 is, in fact, the only SMC line of sight known to exhibit a Milky Way-like extinction curve, with a 2175$\,\AA$ extinction bump \citep{Gordon2003,Welty2010}. In the LMC, Sk\,6619 fits a Milky Way-like extinction curve \citep{Gordon2003} and is marked with a green circle point above the 1:1 line seen in Fig. \ref{fig:av_all}. 

The good agreement of the two extinction estimates for GRB-DLAs can be explained by the fact the two measurements may be probing dust with more similar composition - because there may be less C-rich dust given the low metallicity - and in more similar regions - because the GRB is expected to photo-dissociate the surrounding dense region, \citep[e.g.,][]{Ledoux2009}, possibly leading to $A_{V,\rm ext}$ being more representative of the diffuse warm neutral medium in the host galaxy. In general, the measurements of $A_{V,\rm ext}$ in GRB afterglows are reliable \citep{Watson2006,zafar2013,Perley2013, Heintz2019b}, because the intrinsic afterglow spectrum is known to be either a power law or a broken power law \citep[e.g.][]{Sari1998}.

Overall, $A_{V,\rm depl}$ and $A_{V,\rm ext}$ do not measure dust in the same way or in the same regions. $A_{V,\rm depl}$ measures the average extinction along the line of sight, while only dusty clouds contribute to $A_{V,\rm ext}$ (i.e., the latter is not sensitive to the dust-free segments of the line of sight). The analysis of the depletion is mostly probing the warm neutral medium, and less sensitive to the dense clumps of cold molecular gas, which may contain large fractions of C-rich dust. This is a likely cause of the discrepancy between $A_{V,\rm depl}$ and $A_{V,\rm ext}$ observed in the Milky Way and a few lines of sight in the Magellanic Clouds.

\section{Conclusions}
\label{sec: conclusions}
In this paper we measure the dust-to-metal (DTM), dust-to-gas (DTG) ratios, dust extinction from depletion $A_{V, \rm depl}$ and the fraction of dust mass contributed by element X (f$_{\rm{M}_{X}}$) from the dust depletion of all the elements that contribute to dust. We do this for systems in different environments, QSO-DLAs, GRB-DLAs, LMC, SMC and the Milky Way. We study the DTM, DTG and $A_{V, \rm depl}$ evolution with the total dust-corrected metallicity and the dust tracer [Zn/Fe]. The evolution of the DTM and DTG with metallicity gives us clues on the dominant mechanisms of dust production. We find that the DTM and DTG ratios increase with metallicity and with [Zn/Fe], which suggests that the dominant mechanism for dust production in the metallicity range (-2 $\leq$ [M/H]$_{\rm{tot}}$ $\leq$ 0.5) and redshift range ($0.6 < z < 6.3$) that we study is grain growth in the ISM. Our data are in very good agreement with the dust production and evolution model from \citet{Mattsson2020}. The comparison of our data with the model indicates that the stellar dust yield is only 1$\%$ of the metals yield, which suggests that the net amount of dust that the overall stellar population produces is about ten times less than what was previously estimated. However, we note that this could be an effect of observational bias, since our data probe sightlines of warm diffuse gas and, as a result, might be missing some dust production contributions from stellar sources.

We investigate the cosmic metallicity evolution of the neutral gas and find a declining evolution with increasing redshift, which is probed both with QSO- and GRB-DLAs in a similar way.

The main elements that contribute to dust are, in order of abundance in mass, O, Fe, Si, Mg, C, S, Ni and Al for different environments and with overall similar percentages. Si, Mg and C have similar contributions by mass in dust in all the environments and vary with the amount of dust in each system, as traced by [Zn/Fe]. O and Fe have the highest fractions among all studied elements and show opposite trends with [Zn/Fe] between each other. O (and C) is increasing, while Fe is decreasing with [Zn/Fe]. This opposite evolution may be due to larger availability of Fe in systems with lower levels of depletions. In the low-dust regime ([Zn/Fe] $<$ 0.3), we find that the amount of O in dust is about three times that of Si. This suggests the presence of pyroxenes as the main dust species.

We calculate the dust-to-metal ratio in terms of column density (DTM$_{\rm N}$), in order to estimate the dust extinction using dust depletion ($A_{V,\rm depl}$). The dust extinction $A_{V,\rm depl}$ is increasing with metallicity and with [Zn/Fe]. At any given metallicity, we find lower values of $A_{V,\rm depl}$ for QSO-DLAs than for GRB-DLAs, Mikly Way, and Magellanic Clouds. This suggest that not only metallicity but also the presence of dense cold gas at high pressure (such as in galaxy disks) are determinant factors to facilitate the growth of dust in the ISM.

We compare the dust extinction estimated from depletion ($A_{V,\rm depl}$) with the one estimated from SED or from the reddening ($A_{V,\rm ext}$). We find an overall good agreement between $A_{V,\rm depl}$ and $A_{V,\rm ext}$. However, we find that $A_{V,\rm depl}$ is much lower than $A_{V,\rm ext}$ for the Milky Way, as well as in a few systems in the SMC and LMC. In all these cases, the systems show a 2175$\,\AA$ bump in their extinction curves. The likely cause of the discrepancy is the presence of C-rich dust in clumps of dense cold gas, such as molecular clouds, which are not probed by the depletion measurements. Overall, $A_{V,\rm depl}$ and $A_{V,\rm ext}$ do not measure the amount of dust in the same way and in the same regions, with $A_{V,\rm depl}$ measuring an average extinction along the line of sight and $A_{V,\rm ext}$ being sensitive to dust-rich clouds. Nevertheless, in this work we refine the measurements of $A_{V,\rm depl}$ and find a better consistency with $A_{V,\rm ext}$. The reasons for the discrepancies that remain are probably related to C-rich dust in clumps of cold dense gas.

\begin{acknowledgements}
We thank the anonymous referee for the useful and constructive comments that improved this manuscript. CK, ADC and TRH acknowledge support by the Swiss National Science Foundation under grant 185692. KEH acknowledges support from the Carlsberg Foundation Reintegration Fellowship Grant CF21-0103. ACA would like to acknowledge support by the Carlsberg Foundation. The Cosmic Dawn Center (DAWN) is funded by the Danish National Research Foundation under grant No. 140. This research has made use of NASA’s Astrophysics Data System.
\end{acknowledgements}

\bibliographystyle{aa}
\bibliography{ref}


\appendix

\section{Additional figures}
\label{appsec: dustcomp_all}

Here we present additional figures for the relation between the DTM by mass and the DTM by column density (DTM$_{\rm N}$) and for the dust composition f$_{\rm M_{X}}$ with respect to [Zn/Fe] for all the elements that contribute to dust with f$_{\rm M_{X}}$ > 1$\%$, namely O, S, Si, Mg, Fe, Ni, Al and C. In the dust composition figgures, we show the targets that have dust depletion measurements with filled symbols and the ones whose dust depletion is calculated from the coefficients listed in Table \ref{coefficients} with empty symbols. This is shown for each environment: QSO- and GRB-DLAs, the Milky Way, the LMC and the SMC.

\begin{figure}[!ht]
    \includegraphics[width=\columnwidth]{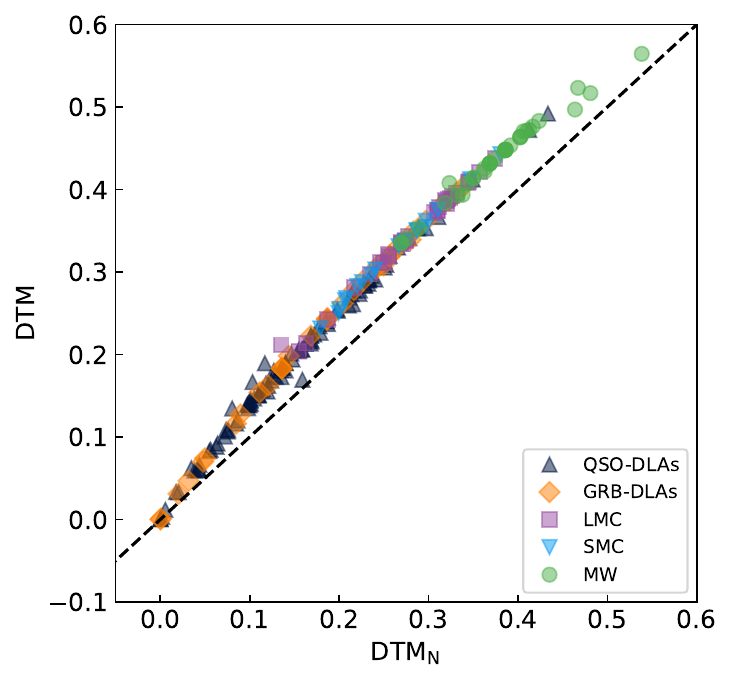}
    \centering
    \caption{Relation between the DTM calculated in terms of mass and the DTM calculated in terms of column density (DTM$_{\rm N}$) for QSO- and GRB-DLAs, LMC, SMC and the Milky Way. The dashed line is the 1:1 line of equality. DTM is on average $\sim$ 0.05 times higher than DTM$_{\rm N}$.}
    \label{fig:dtm-dtmN}
\end{figure}

\begin{figure*}[h!]
\centering
\includegraphics[width=0.8\textwidth]{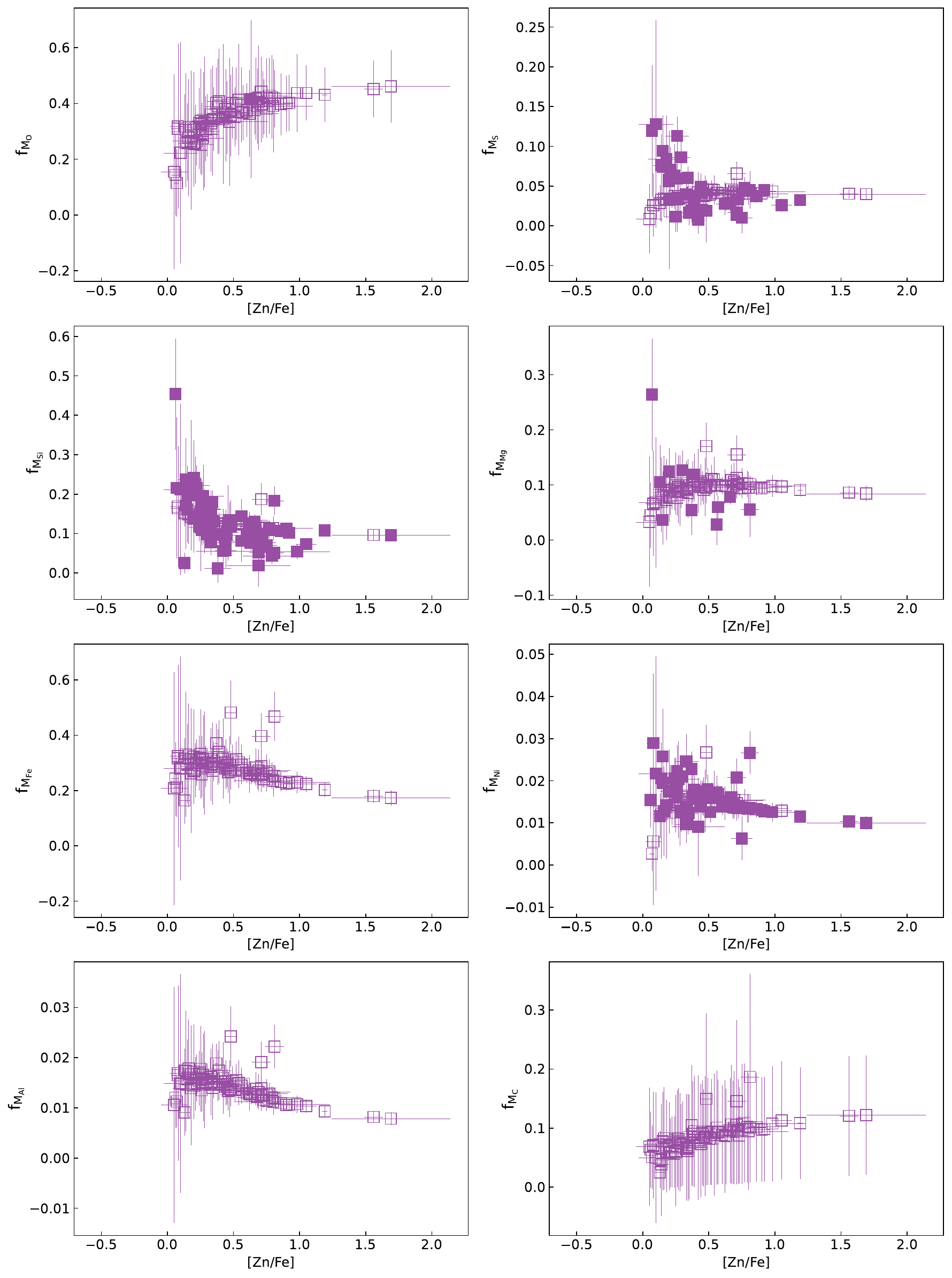}
\caption{Fraction of dust mass contributed by element X, f$_{\rm{M}_{X}}$ as a function of the dust tracer [Zn/Fe] for QSO-DLAs. The filled squares indicate the targets that have a dust depletion measurement and the empty symbols indicate the targets whose dust depletion is calculated from the coefficients listed in Table \ref{coefficients}.}
\label{fig:dustcomp_all_qso_dla}
\end{figure*}

\begin{figure*}[h!]
\centering
\includegraphics[width=0.8\textwidth]{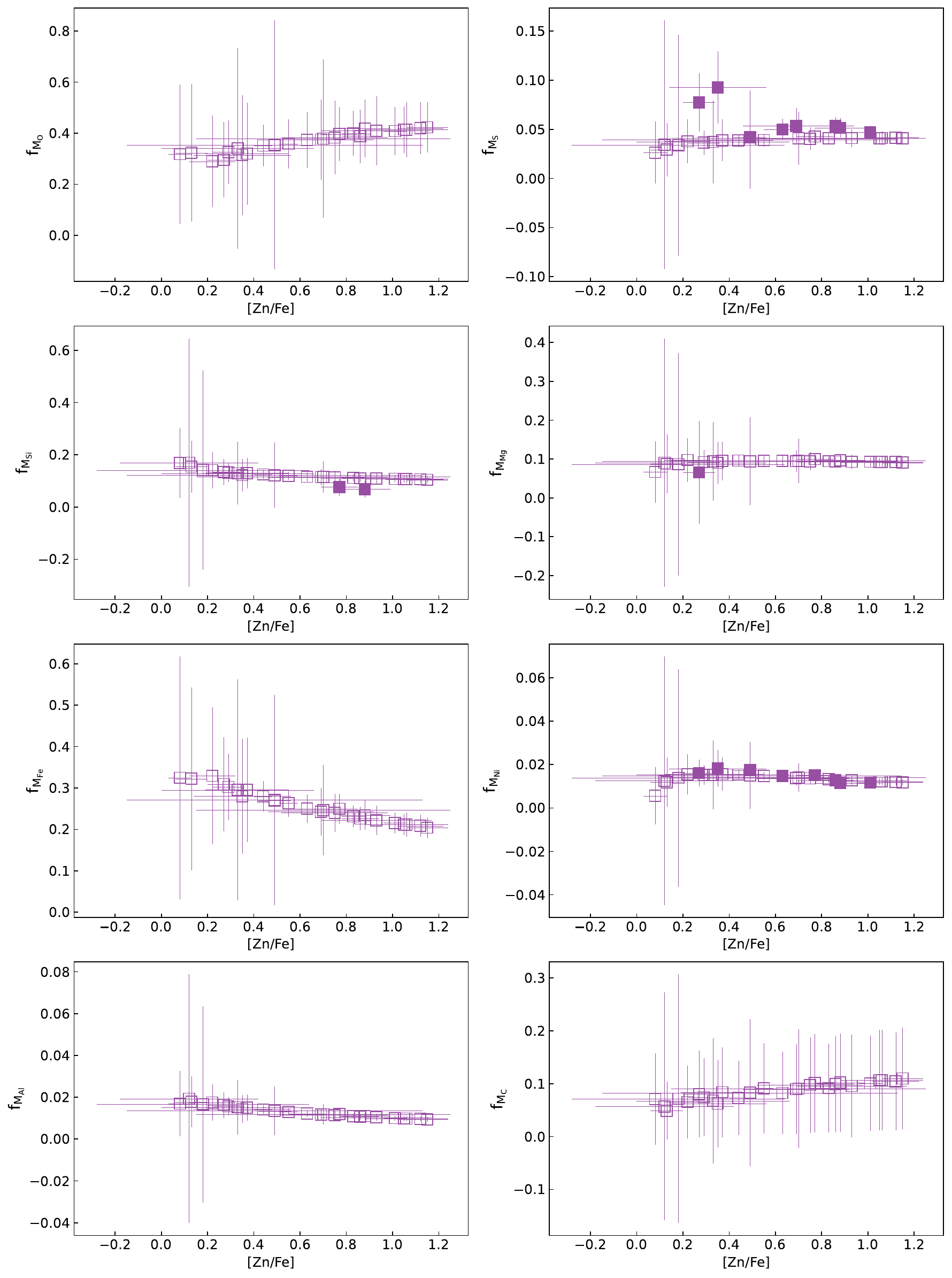}
\caption{Fraction of dust mass contributed by element X, f$_{\rm{M}_{X}}$ as a function of the dust tracer [Zn/Fe] for GRB-DLAs.}
\label{fig:dustcomp_all_grb_dla}
\end{figure*}

\begin{figure*}[h!]
\centering
\includegraphics[width=0.8\textwidth]{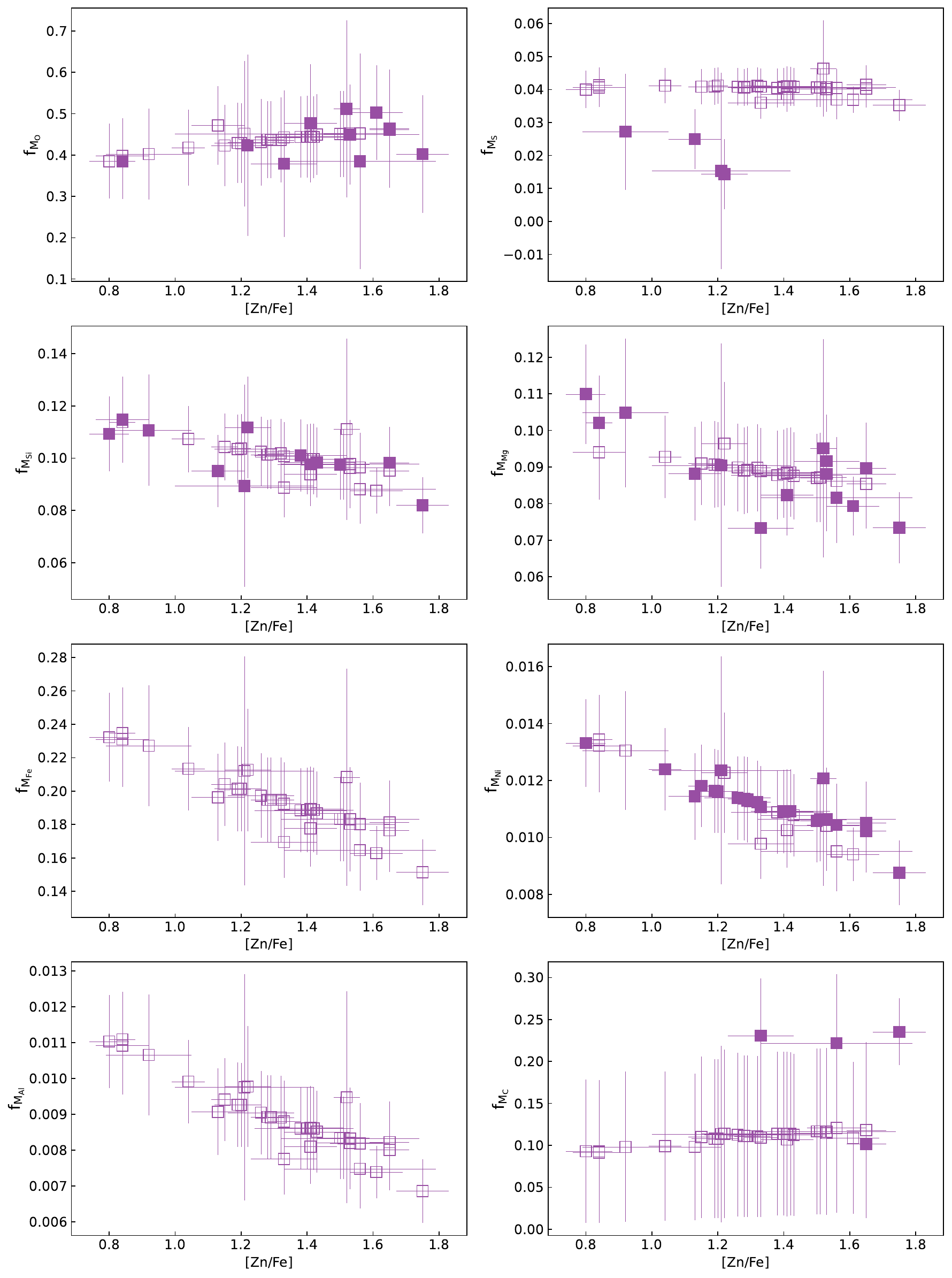}
\caption{Fraction of dust mass contributed by element X, f$_{\rm{M}_{X}}$ as a function of the dust tracer [Zn/Fe] for the Milky Way.}
\label{fig:dustcomp_all_mw}
\end{figure*}

\begin{figure*}[h!]
\centering
\includegraphics[width=0.8\textwidth]{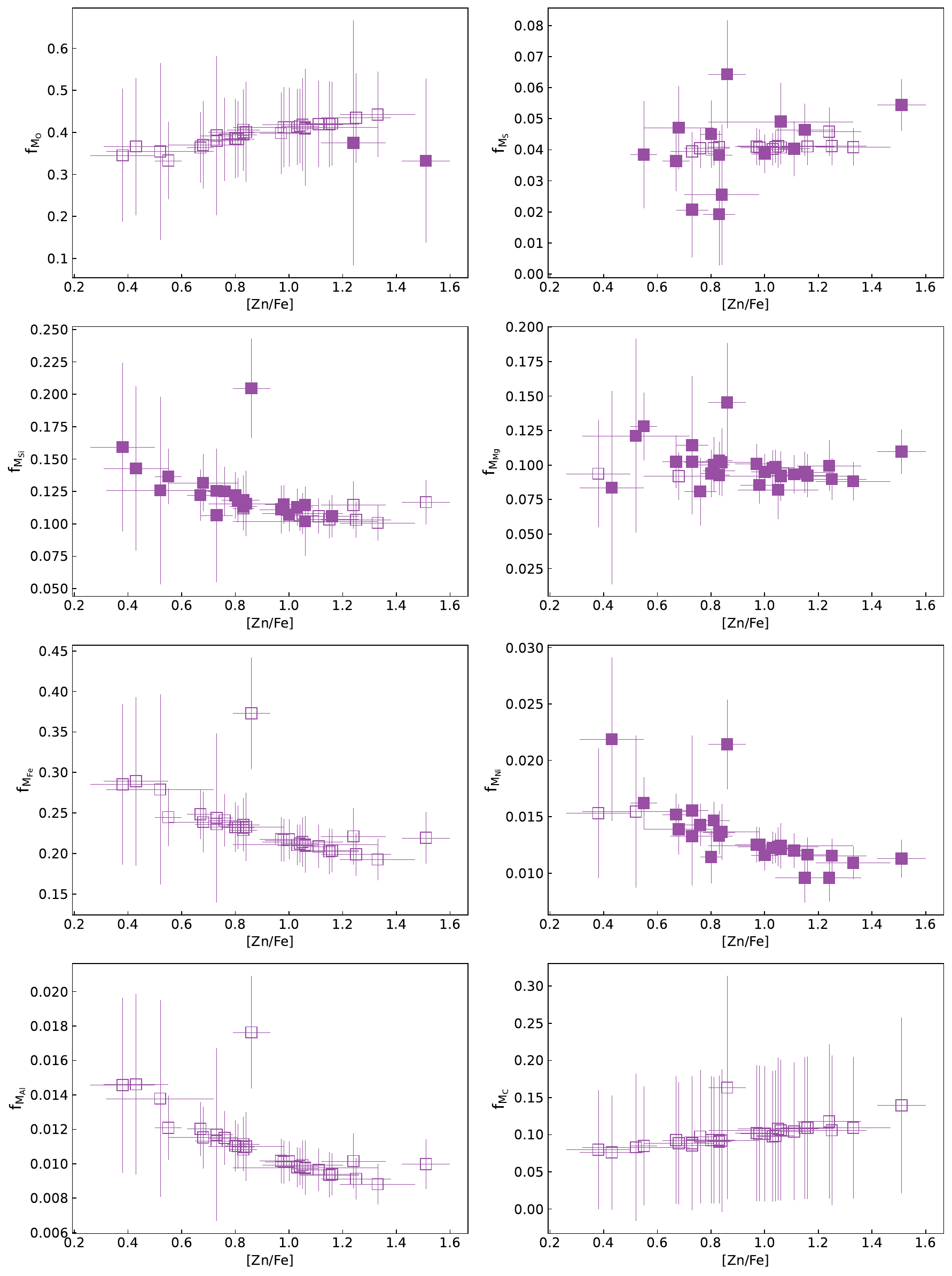}
\caption{Fraction of dust mass contributed by element X, f$_{\rm{M}_{X}}$ as a function of the dust tracer [Zn/Fe] for the LMC.}
\label{fig:dustcomp_all_lmc}
\end{figure*}

\begin{figure*}[h!]
\centering
\includegraphics[width=0.8\textwidth]{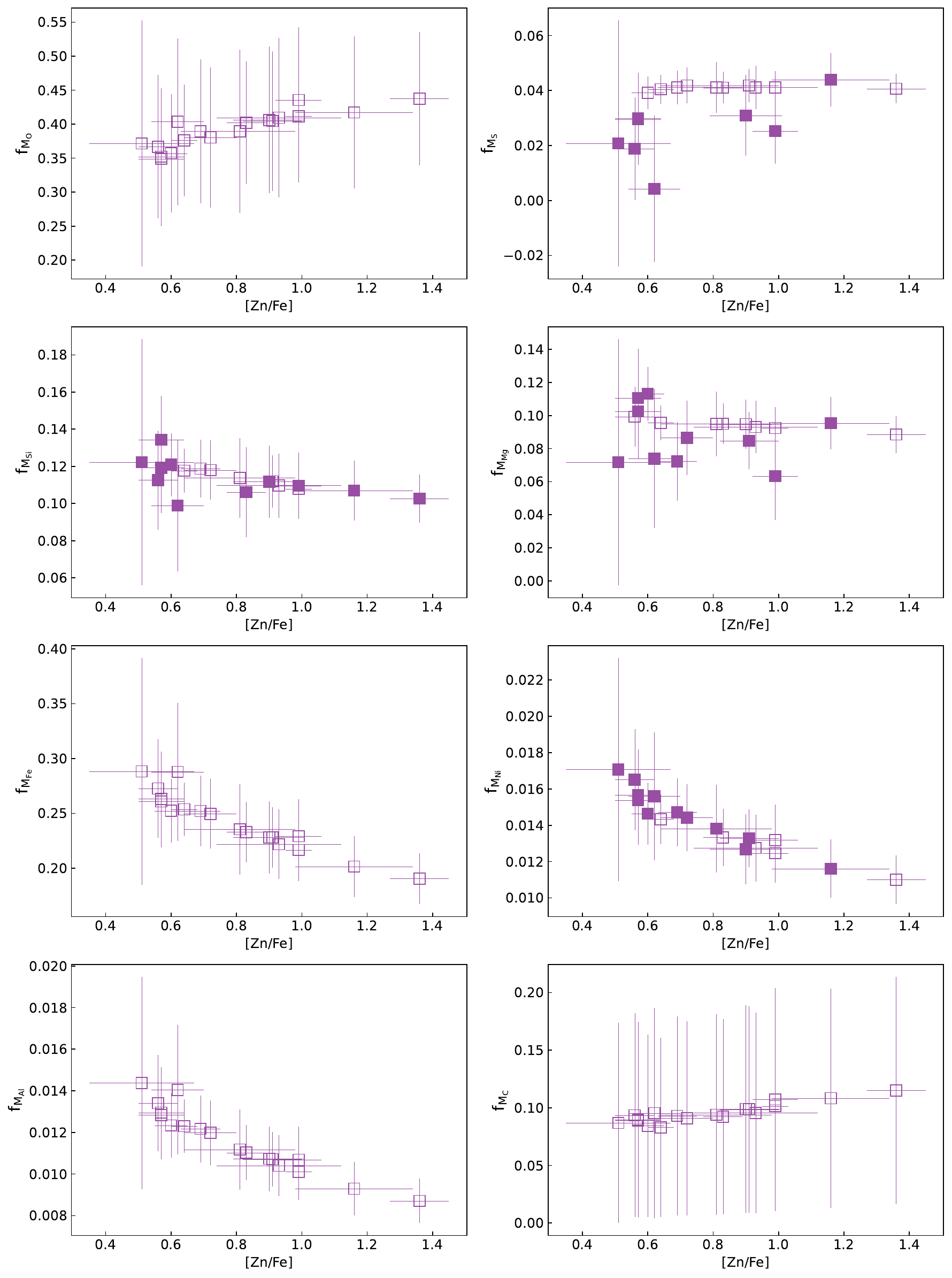}
\caption{Fraction of dust mass contributed by element X, f$_{\rm{M}_{X}}$ as a function of the dust tracer [Zn/Fe] for the SMC.}
\label{fig:dustcomp_all_smc}
\end{figure*}

\clearpage

\clearpage
\section{Dust properties of the data sample}

Here we present the tables of the dust properties for QSO- and GRB-DLAs, the Milky Way, LMC and SMC. We include the redshift z$_{\rm abs}$, the column densities of \ion{H}{i}, \ion{H}{ii}, the dust tracer [Zn/Fe], the DTM and DTG in terms of mass, the DTM in terms of column density (DTM$_{\rm N}$), the dust extinction estimated from the dust depletion $A_{V, \rm depl}$ and the total dust-corrected metallicity [\textit{M}/$\rm{H}$]$_{\rm{tot}}$. We note that the DTM in terms of column density (DTM$_{\rm N}$) is not normalized by the Galactic DTM. We also report the references of the data samples.

\onecolumn
{
\setlength\tabcolsep{1pt}
\begin{longtable}{l@{\hspace{0.2mm}}lcclll@{\hspace{0.2mm}}l@{\hspace{0.2mm}}l@{\hspace{0.2mm}}lc}
\caption{QSO-DLAs dust properties} 
\label{dlastable}\\
\hline \hline
QSO & z$_{\rm abs}$ & log N(\ion{H}{I})& log N(\ion{H}{2}) & [Zn/Fe] & 
DTM & 
DTG & DTM$_{\rm N}$ &
$A_{V, \rm depl}$ &
[M/H]$_{\rm tot}$ & Ref\\
\hline
\noalign{\smallskip}\hline\hline\noalign{\smallskip}
\endfirsthead
\caption{continued}\\
\hline
QSO & z$_{\rm abs}$ & log N(\ion{H}{I})& log N(\ion{H}{2}) & [Zn/Fe] & 
DTM & 
DTG &
DTM$_{\rm N}$ &
$A_{V, \rm depl}$ &
[M/H]$_{\rm tot}$& Ref\\
\noalign{\smallskip}\hline\hline\noalign{\smallskip}
\endhead
\hline
\endfoot
\hline
\endlastfoot
B0105-008      & 1.371  & 21.70$\pm$0.15  & ...            & 0.08$\pm$0.05  & 0.06$\pm$0.02    & 4.00e-05$\pm$2.00e-05 & 0.04$\pm$0.01 & 0.01$\pm$0.01 & -1.35$\pm$0.17 & *,2,3 \\
B2314-409      & 1.857  & 20.90$\pm$0.10   & ...            & 0.18$\pm$0.14  & 0.14$\pm$0.04    & 2.30e-04$\pm$1.20e-04 & 0.10$\pm$0.03  & 0.01$\pm$0.01 & -0.92$\pm$0.19 & *,2,3 \\
B2355-106      & 1.173  & 21.00$\pm$0.10   & ...            & 0.42$\pm$0.20   & 0.19$\pm$0.05    & 5.10e-04$\pm$3.00e-04 & 0.15$\pm$0.05 & 0.04$\pm$0.02 & -0.72$\pm$0.23 & *,2,3 \\
BRI1013+0035   & 3.104   & 21.10$\pm$0.10   & ...            & 0.90$\pm$0.05   & 0.35$\pm$0.04    & 3.82e-03$\pm$1.24e-03 & 0.29$\pm$0.06 & 0.36$\pm$0.15 & -0.11$\pm$0.13 & *,2,3 \\
FJ0812+32      & 2.626  & 21.35$\pm$0.10  & ...            & 0.81$\pm$0.07  & 0.17$\pm$0.14    & 6.10e-04$\pm$5.40e-04 & 0.10$\pm$0.15  & 0.08$\pm$0.12 & -0.58$\pm$0.14 & *,2,3 \\
FJ0812+32      & 2.067  & 21.00$\pm$0.10   & ...            & 0.06$\pm$0.03  & 0.07$\pm$0.02    & 4.00e-05$\pm$1.00e-05 & 0.04$\pm$0.01 & ...           & -1.37$\pm$0.13 & *,2,3 \\
FJ2334-0908    & 3.057  & 20.48$\pm$0.05 & ...            & 0.48$\pm$0.04  & 0.13$\pm$0.15    & 2.90e-04$\pm$3.20e-04 & 0.08$\pm$0.16 & 0.003$\pm$0.001  & -0.81$\pm$0.09 & *,2,3 \\
HE0515-4414    & 1.150  & 20.45$\pm$0.15 & ...            & 0.54$\pm$0.20   & 0.23$\pm$0.05    & 5.10e-04$\pm$3.20e-04 & 0.18$\pm$0.05 & 0.01$\pm$0.01 & -0.79$\pm$0.26 & *,2,3 \\
HE1104-1805    & 1.662  & 20.85$\pm$0.01 & ...            & 0.46$\pm$0.02  & 0.39$\pm$0.10     & 7.80e-04$\pm$2.30e-04 & 0.35$\pm$0.10 & 0.05$\pm$0.02 & -0.84$\pm$0.07 & *,2,3 \\
J0000+0048     & 2.525  & 21.07$\pm$0.10  & ...            & 1.69$\pm$0.45  & 0.49$\pm$0.07    & 5.43e-02$\pm$6.06e-02 & 0.43$\pm$0.09 & 5.14$\pm$5.91 & 0.90$\pm$0.48   & *,2,3 \\
J0008-0958     & 1.768  & 20.85$\pm$0.15 & ...            & 0.44$\pm$0.07  & 0.21$\pm$0.03    & ... & 0.16$\pm$0.03 & ...           & 0.0$\pm$0.18   & *,2,3 \\
J0058+0115     & 2.010  & 21.10$\pm$0.15  & ...            & 0.51$\pm$0.07  & 0.24$\pm$0.03    & 8.50e-04$\pm$3.70e-04 & 0.19$\pm$0.04 & 0.08$\pm$0.04 & -0.60$\pm$0.18  & *,2,3 \\
J0256+0110     & 0.725  & 20.70$\pm$0.16  & ...            & 0.80$\pm$0.30    & 0.33$\pm$0.06    & 6.18e-03$\pm$5.38e-03 & 0.27$\pm$0.07 & 0.23$\pm$0.22 & 0.13$\pm$0.37  & *,2,3 \\
J0927+1543     & 1.731  & ...            & ...            & 0.98$\pm$0.25  & 0.35$\pm$0.06    & ... & 0.30$\pm$0.07  & ...           & ...            & *,2,3 \\
J1049-0110     & 1.658  & 20.31$\pm$0.15 & ...            & 0.71$\pm$0.07  & 0.26$\pm$0.04    & 9.32e-03$\pm$4.11e-03 & 0.22$\pm$0.05 & 0.14$\pm$0.08 & 0.41$\pm$0.18  & *,2,3 \\
J1056+1208     & 1.610  & 21.45$\pm$0.15 & ...            & 0.69$\pm$0.07  & 0.28$\pm$0.04    & 3.20e-03$\pm$1.39e-03 & 0.23$\pm$0.05 & 0.67$\pm$0.39 & -0.09$\pm$0.18 & *,2,3 \\
J1107+0048     & 0.740  & 21.00$\pm$0.05  & ...            & 0.28$\pm$0.105  & 0.16$\pm$0.03    & 8.00e-04$\pm$3.50e-04 & 0.12$\pm$0.03 & 0.05$\pm$0.03 & -0.45$\pm$0.17 & *,2,3 \\
J1135-0010     & 2.207  & 22.05$\pm$0.10  & ...            & 0.60$\pm$0.04   & 0.27$\pm$0.03    & 5.20e-04$\pm$1.70e-04 & 0.21$\pm$0.04 & 0.42$\pm$0.18 & -0.86$\pm$0.13 & *,2,3 \\
J1142+0701     & 1.841  & 21.50$\pm$0.15  & ...            & 0.56$\pm$0.07  & 0.25$\pm$0.03    & 7.90e-04$\pm$3.40e-04 & 0.20$\pm$0.04  & 0.18$\pm$0.10  & -0.65$\pm$0.18 & *,2,3 \\
J1155+0530     & 3.326  & 21.05$\pm$0.10  & ...            & 0.26$\pm$0.09  & 0.18$\pm$0.02    & 5.20e-04$\pm$1.90e-04 & 0.13$\pm$0.03 & 0.04$\pm$0.02 & -0.68$\pm$0.15 & *,2,3 \\
J1200+4015     & 3.220  & 20.65$\pm$0.15 & ...            & 0.29$\pm$0.07  & 0.17$\pm$0.02    & 1.22e-03$\pm$5.30e-04 & 0.13$\pm$0.03 & 0.04$\pm$0.02 & -0.30$\pm$0.18  & *,2,3 \\
J1211+0833     & 2.117  & 21.00$\pm$0.20   & ...            & 1.56$\pm$0.07  & 0.47$\pm$0.06    & 1.73e-02$\pm$9.02e-03 & 0.41$\pm$0.08 & 1.38$\pm$0.98 & 0.42$\pm$0.22  & *,2,3 \\
J1237+0647     & 2.690  & 20.00$\pm$0.15  & ...            & 1.19$\pm$0.02  & 0.41$\pm$0.05    & 3.37e-02$\pm$1.38e-02 & 0.35$\pm$0.07 & 0.26$\pm$0.15 & 0.77$\pm$0.17  & *,2,3 \\
J1240+1455     & 3.108  & 21.30$\pm$0.20   & ...            & 1.05$\pm$0.08  & 0.37$\pm$0.05    & 9.90e-04$\pm$5.40e-04 & 0.31$\pm$0.06 & 0.15$\pm$0.11 & -0.71$\pm$0.23 & *,2,3 \\
J1310+5424     & 1.801   & 21.45$\pm$0.15 & ...            & 0.67$\pm$0.07  & 0.26$\pm$0.05    & 1.85e-03$\pm$8.30e-04 & 0.21$\pm$0.05 & 0.39$\pm$0.23 & -0.29$\pm$0.18 & *,2,3 \\
J1313+1441     & 1.795   & ...            & ...            & 0.49$\pm$0.07  & 0.21$\pm$0.03    & ... & 0.17$\pm$0.04 & ...           & ...            & *,2,3 \\
J1417+4132     & 1.951  & 21.45$\pm$0.25 & ...            & 0.71$\pm$0.07  & 0.19$\pm$0.19    & ... & 0.12$\pm$0.21 & ...           & ...            & *,2,3 \\
J1431+3952     & 0.602   & 21.20$\pm$0.10   & ...            & 0.62$\pm$0.22  & 0.28$\pm$0.05    & 9.90e-04$\pm$5.90e-04 & 0.22$\pm$0.05 & 0.11$\pm$0.08 & -0.59$\pm$0.25 & *,2,3 \\
J1454+0941     & 1.788  & 20.50$\pm$0.15  & ...            & 0.44$\pm$0.13  & 0.21$\pm$0.03    & 1.65e-03$\pm$8.40e-04 & 0.16$\pm$0.04 & 0.04$\pm$0.02 & -0.24$\pm$0.21 & *,2 \\
J1541+3153     & 2.444  & 20.95$\pm$0.10  & ...            & 0.27$\pm$0.16  & 0.16$\pm$0.03    & 8.00e-05$\pm$4.00e-05 & 0.12$\pm$0.03 & ...           & -1.45$\pm$0.20  & *,2,3 \\
J1552+4910     & 1.960  & ...            & ...            & 0.20$\pm$0.07   & 0.12$\pm$0.03    & 6.00e-05$\pm$2.00e-05 & 0.09$\pm$0.02 & ...           & -1.46$\pm$0.13 & *,2,3 \\
J1604+3951     & 3.163  & 21.75$\pm$0.20  & ...            & 0.39$\pm$0.07  & 0.20$\pm$0.02     & 2.10e-04$\pm$1.10e-04 & 0.15$\pm$0.03 & 0.08$\pm$0.06 & -1.12$\pm$0.22 & *,2,3 \\
J1623+0718     & 1.336  & 21.35$\pm$0.10  & ...            & 0.38$\pm$0.10   & 0.18$\pm$0.03    & 3.00e-04$\pm$1.20e-04 & 0.14$\pm$0.03 & 0.05$\pm$0.02 & -0.93$\pm$0.16 & *,2,3 \\
J2328+0022     & 0.652  & 20.32$\pm$0.07 & ...            & 0.34$\pm$0.15  & 0.18$\pm$0.03    & 1.04e-03$\pm$4.70e-04 & 0.14$\pm$0.04 & 0.01$\pm$0.01 & -0.39$\pm$0.18 & *,2,3 \\
J2340-00       & 2.054  & 20.35$\pm$0.15 & ...            & 0.39$\pm$0.09  & 0.17$\pm$0.03    & 1.51e-03$\pm$7.20e-04 & 0.14$\pm$0.04 & 0.02$\pm$0.01 & -0.20$\pm$0.19  & *,2,3 \\
PSS1253-0228   & 2.783  & 21.85$\pm$0.20  & ...            & 0.16$\pm$0.08  & 0.10$\pm$0.02     & 3.00e-05$\pm$2.00e-05 & 0.07$\pm$0.02 & 0.01$\pm$0.01 & -1.64$\pm$0.23 & *,2,3 \\
PSSJ2344+0342  & 3.220  & 21.25$\pm$0.08 & ...            & -0.08$\pm$0.34 & 0.0011$\pm$0.120  & 3.00e-07$\pm$4.00e-05 & 0.001 
 $\pm$0.08  & ...           & -1.65$\pm$0.35 & *,2,3 \\
Q0000-263      & 3.390  & 21.40$\pm$0.08  & ...            & -0.11$\pm$0.06 & 0.003$\pm$0.16   & 3.00e-07$\pm$2.00e-05 & 0.003$\pm$0.16  & ...           & -2.14$\pm$0.11 & *,1,3 \\
Q0010-002      & 2.025 & 20.95$\pm$0.10  & ...            & -0.25$\pm$0.06 & 0.0005$\pm$0.030  & 2.00e-07$\pm$1.00e-05 & 0.001$\pm$0.03  & ...           & -1.49$\pm$0.13 & *,1,3 \\
Q0013-004      & 1.973 & 20.83$\pm$0.05 & 18.86$\pm$1.14 & 0.75$\pm$0.01   & 0.32$\pm$0.04    & 1.77e-03$\pm$4.60e-04 & 0.26$\pm$0.05 & 0.09$\pm$0.03 & -0.40$\pm$0.10   & *,1,3 \\
Q0027-1836     & 2.402   & 21.75$\pm$0.10  & 17.30$\pm$0.07  & 0.56$\pm$0.03  & 0.24$\pm$0.03    & 1.30e-04$\pm$4.00e-05 & 0.19$\pm$0.04 & 0.05$\pm$0.02 & -1.41$\pm$0.13 & *,2,3 \\
Q0058-292      & 2.671 & 21.10$\pm$0.10   & ...            & 0.24$\pm$0.03  & 0.14$\pm$0.01    & 6.00e-05$\pm$2.00e-05 & 0.10$\pm$0.02  & 0.01$\pm$0.001 & -1.48$\pm$0.11 & *,1,3 \\
Q0100+130      & 2.309 & 21.35$\pm$0.08 & ...            & 0.13$\pm$0.01  & 0.09$\pm$0.01    & 3.00e-05$\pm$1.00e-05 & 0.06$\pm$0.01 & ...           & -1.62$\pm$0.11 & *,1,3 \\
Q0102-190a     & 2.370  & 21.00$\pm$0.08  & ...            & 0.01$\pm$0.08  & 0.01$\pm$0.03    & ... & 0.01$\pm$0.02 & ...           & ...            & *,1,3 \\
Q0149+33       & 2.141  & 20.50$\pm$0.10   & ...            & 0.05$\pm$0.10   & 0.06$\pm$0.03    & 2.00e-05$\pm$1.00e-05 & 0.04$\pm$0.02 & ...           & -1.59$\pm$0.16 & *,2,3 \\
Q0201+365      & 2.463  & 20.38$\pm$0.05 & ...            & 0.20$\pm$0.05   & 0.14$\pm$0.02    & 1.67e-03$\pm$4.70e-04 & 0.10$\pm$0.02  & 0.03$\pm$0.01 & -0.07$\pm$0.11 & *,2,3 \\
Q0216+080a     & 1.769 & 20.30$\pm$0.10   & ...            & 0.24$\pm$0.06  & 0.15$\pm$0.02    & 1.90e-04$\pm$7.00e-05 & 0.1$\pm$0.02  & ...           & -1.02$\pm$0.13 & *,1,3 \\
Q0216+080b     & 2.293 & 20.50$\pm$0.10   & ...            & 0.33$\pm$0.05  & 0.17$\pm$0.02    & 5.90e-04$\pm$2.20e-04 & 0.13$\pm$0.03 & 0.01$\pm$0.01 & -0.61$\pm$0.15 & *,1,3 \\
Q0302-223      & 1.009  & 20.36$\pm$0.11 & ...            & 0.52$\pm$0.08  & 0.22$\pm$0.03    & 1.31e-03$\pm$4.90e-04 & 0.17$\pm$0.04 & 0.02$\pm$0.01 & -0.36$\pm$0.15 & *,2,3 \\
Q0347-383      & 3.025 & 20.73$\pm$0.05 & 14.53$\pm$0.06 & 0.63$\pm$0.05  & 0.28$\pm$0.08    & 3.10e-04$\pm$1.50e-04 & 0.22$\pm$0.09 & 0.01$\pm$0.01 & -1.10    $\pm$0.18  & *,1,3 \\
Q0405-443a     & 1.913 & 20.80$\pm$0.10   & ...            & 0.02$\pm$0.04  & 0.03$\pm$0.02    & 4.00e-05$\pm$2.00e-05 & 0.02$\pm$0.01 & ...           & -1.08$\pm$0.12 & *,1,3 \\
Q0405-443b     & 2.550  & 21.15$\pm$0.15 & ...            & 0.15$\pm$0.04  & 0.11$\pm$0.02    & 5.00e-05$\pm$2.00e-05 & 0.08$\pm$0.02 & ...           & -1.46$\pm$0.18 & *,1,3 \\
Q0405-443c     & 2.595 & 21.05$\pm$0.10  & 18.14$\pm$0.07 & 0.26$\pm$0.01  & 0.15$\pm$0.02    & 1.50e-04$\pm$4.00e-05 & 0.11$\pm$0.02 & 0.01$\pm$0.001  & -1.14$\pm$0.12 & *,1,3 \\
Q0449-1645     & 1.007  & 20.98$\pm$0.07 & ...            & 0.27$\pm$0.07  & 0.16$\pm$0.02    & 2.90e-04$\pm$9.00e-05 & 0.11$\pm$0.03 & 0.02$\pm$0.01 & -0.88$\pm$0.12 & *,2,3 \\
Q0458-020      & 2.040  & 21.70$\pm$0.10   & ...            & 0.57$\pm$0.04  & 0.25$\pm$0.03    & 2.50e-04$\pm$1.00e-04 & 0.2$\pm$0.04  & 0.09$\pm$0.04 & -1.15$\pm$0.16 & *,1,3 \\
Q0528-250a     & 2.141 & 20.98$\pm$0.05 & ...            & 0.25$\pm$0.04  & 0.14$\pm$0.02    & 9.00e-05$\pm$2.00e-05 & 0.1$\pm$0.02  & 0.010$\pm$0.001  & -1.35$\pm$0.08 & *,1,3 \\
Q0528-250b     & 2.811 & 21.35$\pm$0.07 & 18.22$\pm$0.11 & 0.37$\pm$0.01  & 0.16$\pm$0.02    & 2.70e-04$\pm$9.00e-05 & 0.12$\pm$0.03 & 0.04$\pm$0.02 & -0.90$\pm$0.12  & *,1,3 \\
Q0551-366      & 1.962 & 20.70$\pm$0.08  & 17.42$\pm$0.14 & 0.71$\pm$0.04  & 0.29$\pm$0.04    & 2.26e-03$\pm$6.50e-04 & 0.23$\pm$0.05 & 0.08$\pm$0.03 & -0.25$\pm$0.11 & *,1,3 \\
Q0642-5038     & 2.659  & 20.95$\pm$0.08 & ...            & 0.22$\pm$0.05  & 0.14$\pm$0.02    & 3.00e-05$\pm$1.00e-05 & 0.10$\pm$0.02  & ...           & -1.83$\pm$0.12 & *,2,3 \\
Q0841+129a     & 1.864 & 21.00$\pm$0.10   & ...            & -0.10$\pm$0.12  & 0.00017$\pm$0.04 & ... & 0.0002$\pm$0.03  & ...           & ...            & *,1 \\
Q0841+129b     & 2.375 & 21.05$\pm$0.10  & ...            & 0.15$\pm$0.03  & 0.11$\pm$0.02    & 4.00e-05$\pm$1.00e-05 & 0.08$\pm$0.02 & ...           & -1.61$\pm$0.11 & *,1,3 \\
Q0841+129c     & 2.476 & 20.80$\pm$0.10   & ...            & 0.13$\pm$0.04  & 0.17$\pm$0.11    & 5.00e-05$\pm$4.00e-05 & 0.16$\pm$0.12 & ...           & -1.65$\pm$0.14 & *,1,3 \\
Q0918+1636     & 2.583  & 20.96$\pm$0.05 & ...            & 0.72$\pm$0.01  & 0.29$\pm$0.04    & 1.31e-03$\pm$9.50e-04 & 0.23$\pm$0.05 & 0.09$\pm$0.07 & -0.49$\pm$0.31 & *,2,3 \\
Q0918+1636     & 2.412  & 21.26$\pm$0.06 & ...            & 0.47$\pm$0.29  & 0.23$\pm$0.06    & 3.65e-03$\pm$1.16e-03 & 0.18$\pm$0.06 & 0.46$\pm$0.20  & 0.05$\pm$0.09  & *,2,3 \\
Q0933+733      & 1.479  & 21.62$\pm$0.10  & ...            & 0.26$\pm$0.02  & 0.15$\pm$0.02    & 8.00e-05$\pm$2.00e-05 & 0.11$\pm$0.02 & 0.02$\pm$0.01 & -1.44$\pm$0.12 & *,2,3 \\
Q0935+417      & 1.373  & 20.52$\pm$0.10  & ...            & 0.18$\pm$0.10   & 0.12$\pm$0.03    & 2.60e-04$\pm$1.10e-04 & 0.08$\pm$0.02 & 0.01$\pm$0.0  & -0.80$\pm$0.16  & *,2,3 \\
Q0948+433      & 1.233  & 21.62$\pm$0.06 & ...            & 0.34$\pm$0.01  & 0.18$\pm$0.02    & 2.70e-04$\pm$6.00e-05 & 0.14$\pm$0.02 & 0.08$\pm$0.02 & -0.97$\pm$0.09 & *,2,3 \\
Q1010+0003     & 1.265  & 21.52$\pm$0.07 & ...            & 0.44$\pm$0.08  & 0.21$\pm$0.03    & 2.60e-04$\pm$9.00e-05 & 0.16$\pm$0.03 & 0.06$\pm$0.02 & -1.04$\pm$0.13 & *,2,3 \\
Q1037-270      & 2.139 & 19.70$\pm$0.05  & ...            & 0.07$\pm$0.02  & 0.08$\pm$0.02    & ... & 0.05$\pm$0.02 & ...           & ...            & *,1,3 \\
Q1111-152      & 3.266 & 21.30$\pm$0.05  & ...            & 0.25$\pm$0.09  & 0.13$\pm$0.03    & 4.00e-05$\pm$2.00e-05 & 0.10$\pm$0.03  & 0.010$\pm$0.001  & -1.66$\pm$0.14 & *,1,3 \\
Q1117-134      & 3.350  & 20.95$\pm$0.10  & ...            & 0.14$\pm$0.06  & 0.09$\pm$0.02    & 4.00e-05$\pm$1.00e-05 & 0.06$\pm$0.02 & ...           & -1.46$\pm$0.11 & *,1,3 \\
Q1137+3907     & 0.720  & 21.10$\pm$0.10   & ...            & 0.73$\pm$0.07  & 0.31$\pm$0.04    & 3.80e-03$\pm$1.31e-03 & 0.25$\pm$0.05 & 0.35$\pm$0.15 & -0.06$\pm$0.14 & *,2,3 \\
Q1157+014      & 1.944 & 21.80$\pm$0.10   & ...            & 0.30$\pm$0.01   & 0.17$\pm$0.02    & 9.00e-05$\pm$3.00e-05 & 0.12$\pm$0.03 & 0.04$\pm$0.02 & -1.42$\pm$0.11 & *,1,3 \\
Q1209+093      & 2.584 & 21.40$\pm$0.10   & ...            & 0.45$\pm$0.03  & 0.22$\pm$0.03    & 3.20e-04$\pm$1.00e-04 & 0.17$\pm$0.03 & 0.06$\pm$0.02 & -0.98$\pm$0.13 & *,1,3 \\
Q1210+17       & 1.892  & 20.63$\pm$0.08 & ...            & 0.14$\pm$0.06  & 0.11$\pm$0.02    & 2.50e-04$\pm$8.00e-05 & 0.07$\pm$0.02 & 0.010$\pm$0.001  & -0.78$\pm$0.12 & *,2,3 \\
Q1215+33       & 1.999   & 20.95$\pm$0.07 & ...            & 0.33$\pm$0.07  & 0.19$\pm$0.02    & 2.00e-04$\pm$6.00e-05 & 0.14$\pm$0.03 & 0.01$\pm$0.0  & -1.13$\pm$0.12 & *,2,3 \\
Q1223+178      & 2.466 & 21.40$\pm$0.10   & ...            & -0.03$\pm$0.02 & 0.06$\pm$0.01    & 2.00e-05$\pm$1.00e-05 & 0.040$\pm$0.001  & ...           & -1.64$\pm$0.14 & *,1,3 \\
Q1232+082      & 2.338  & 20.90$\pm$0.10   & ...            & 0.77$\pm$0.14  & 0.29$\pm$0.05    & 1.07e-03$\pm$4.70e-04 & 0.23$\pm$0.06 & 0.06$\pm$0.03 & -0.57$\pm$0.18 & *,2,3 \\
Q1328+307      & 0.692  & 21.25$\pm$0.10  & ...            & 0.48$\pm$0.14  & 0.24$\pm$0.04    & 3.30e-04$\pm$1.60e-04 & 0.19$\pm$0.04 & 0.04$\pm$0.02 & -1.00$\pm$0.19  & *,2,3 \\
Q1331+170      & 1.776 & 21.15$\pm$0.07 & ...            & 0.66$\pm$0.04  & 0.29$\pm$0.03    & 2.90e-04$\pm$9.00e-05 & 0.23$\pm$0.04 & 0.03$\pm$0.01 & -1.14$\pm$0.12 & *,1,3 \\
Q1354+258      & 1.420  & 21.54$\pm$0.06 & ...            & 0.33$\pm$0.09  & 0.18$\pm$0.02    & 1.32e-03$\pm$8.40e-04 & 0.14$\pm$0.03 & 0.31$\pm$0.20  & -0.29$\pm$0.27 & *,2,3 \\
Q1409+095a     & 2.019 & 20.65$\pm$0.10  & ...            & 0.10$\pm$0.13   & 0.08$\pm$0.04    & 2.00e-05$\pm$1.00e-05 & 0.06$\pm$0.03 & ...           & -1.70$\pm$0.15  & *,1,3 \\
Q1425+6039     & 2.827  & 20.30$\pm$0.04  & ...            & 0.44$\pm$0.04  & 0.22$\pm$0.02    & 8.00e-04$\pm$1.80e-04 & 0.17$\pm$0.03 & 0.01$\pm$0.0  & -0.59$\pm$0.09 & *,2,3 \\
Q1444+014      & 2.087 & 20.25$\pm$0.07 & 18.16$\pm$0.14 & 0.86$\pm$0.07  & 0.34$\pm$0.05    & 1.11e-03$\pm$5.10e-04 & 0.28$\pm$0.06 & 0.01$\pm$0.01 & -0.63$\pm$0.19 & *,1,3 \\
Q1451+123a     & 2.255 & 20.35$\pm$0.10 & ...            & 0.27$\pm$0.12  & 0.15$\pm$0.03    & 2.10e-04$\pm$9.00e-05 & 0.11$\pm$0.03 & ...           & -1.00$\pm$0.15  & *,1,3 \\
Q1727+5302     & 1.031  & 21.41$\pm$0.15 & ...            & 0.69$\pm$0.24  & 0.27$\pm$0.05    & 1.86e-03$\pm$7.70e-04 & 0.22$\pm$0.06 & 0.36$\pm$0.20  & -0.31$\pm$0.16 & *,2,3 \\
Q1727+5302     & 0.945  & 21.16$\pm$0.10  & ...            & 0.71$\pm$0.11  & 0.29$\pm$0.04    & 3.50e-04$\pm$2.40e-04 & 0.23$\pm$0.05 & 0.04$\pm$0.03 & -1.06$\pm$0.29 & *,2,3 \\
Q1755+578      & 1.971  & 21.40$\pm$0.15  & ...            & 0.81$\pm$0.07  & 0.31$\pm$0.04    & 5.16e-03$\pm$2.24e-03 & 0.25$\pm$0.05 & 0.97$\pm$0.56 & 0.08$\pm$0.18  & *,2,3 \\
Q2116-358      & 1.996 & 20.10$\pm$0.07  & ...            & 0.45$\pm$0.09  & 0.22$\pm$0.03    & 1.67e-03$\pm$4.50e-04 & 0.17$\pm$0.04 & 0.01$\pm$0.01 & -0.27$\pm$0.10  & *,1,3 \\
Q2138-444a     & 2.383 & 20.60$\pm$0.05  & ...            & 0.33$\pm$0.08  & 0.19$\pm$0.02    & 2.20e-04$\pm$6.00e-05 & 0.14$\pm$0.03 & 0.01$\pm$0.0  & -1.07$\pm$0.11 & *,1,3 \\
Q2138-444b     & 2.852 & 20.98$\pm$0.05 & ...            & 0.02$\pm$0.03  & 0.06$\pm$0.01    & 1.00e-05$\pm$0.00e+00 & 0.03$\pm$0.01 & ...           & -1.86$\pm$0.10  & *,1,3 \\
Q2206-199a     & 1.921 & 20.67$\pm$0.05 & ...            & 0.20$\pm$0.01   & 0.14$\pm$0.02    & 6.20e-04$\pm$2.20e-04 & 0.10$\pm$0.02  & 0.02$\pm$0.01 & -0.51$\pm$0.15 & *,1,3 \\
Q2228-3954     & 2.095   & 21.20$\pm$0.10   & ...            & 0.08$\pm$0.06  & 0.06$\pm$0.02    & 5.00e-05$\pm$2.00e-05 & 0.04$\pm$0.01 & ...           & -1.26$\pm$0.14 & *,2,3 \\
Q2230+025      & 1.864  & 20.83$\pm$0.05 & ...            & 0.35$\pm$0.07  & 0.18$\pm$0.02    & 7.60e-04$\pm$2.20e-04 & 0.13$\pm$0.03 & 0.03$\pm$0.01 & -0.52$\pm$0.11 & *,2,3 \\
Q2231-00       & 2.066  & 20.53$\pm$0.08 & ...            & 0.21$\pm$0.06  & 0.15$\pm$0.02    & 3.50e-04$\pm$1.10e-04 & 0.10$\pm$0.02  & 0.01$\pm$0.0  & -0.76$\pm$0.12 & *,2,3 \\
Q2243-605      & 2.331 & 20.65$\pm$0.05 & ...            & 0.26$\pm$0.02  & 0.16$\pm$0.02    & 3.50e-04$\pm$1.10e-04 & 0.11$\pm$0.02 & 0.01$\pm$0.0  & -0.79$\pm$0.12 & *,1,3 \\
Q2318-1107     & 1.989  & 20.68$\pm$0.05 & 15.49$\pm$0.03 & 0.34$\pm$0.03  & 0.20$\pm$0.02     & 5.80e-04$\pm$1.40e-04 & 0.15$\pm$0.02 & 0.02$\pm$0.01 & -0.68$\pm$0.09 & *,2,3 \\
Q2332-094a     & 2.287 & 20.07$\pm$0.07 & ...            & 0.92$\pm$0.04  & 0.35$\pm$0.04    & 3.09e-03$\pm$1.07e-03 & 0.29$\pm$0.06 & 0.03$\pm$0.01 & -0.20$\pm$0.14  & *,1,3 \\
Q2343+125      & 2.431 & 20.40$\pm$0.07  & 13.69$\pm$0.09 & 0.39$\pm$0.03  & 0.21$\pm$0.02    & 3.90e-04$\pm$1.00e-04 & 0.16$\pm$0.03 & 0.01$\pm$0.0  & -0.86$\pm$0.10  & *,1,3 \\
Q2359-022a     & 2.095 & 20.65$\pm$0.10  & ...            & 0.75$\pm$0.08  & 0.29$\pm$0.04    & 1.19e-03$\pm$5.20e-04 & 0.24$\pm$0.05 & 0.04$\pm$0.02 & -0.53$\pm$0.18 & *,1,3 \\
SDSS0225+0054  & 2.714  & 21.00$\pm$0.15  & ...            & 0.33$\pm$0.14  & 0.18$\pm$0.03    & 6.30e-04$\pm$3.20e-04 & 0.14$\pm$0.04 & 0.04$\pm$0.03 & -0.61$\pm$0.21 & *,2,3 \\
SDSS1116+4118A & 2.662  & 20.48$\pm$0.10  & ...            & 0.79$\pm$0.22  & 0.31$\pm$0.05    & 1.47e-03$\pm$8.80e-04 & 0.25$\pm$0.06 & 0.03$\pm$0.02 & -0.46$\pm$0.25 & *,2,3 \\
SDSS1249-0233  & 1.781  & 21.45$\pm$0.15 & ...            & 0.42$\pm$0.05  & 0.21$\pm$0.02    & 4.70e-04$\pm$1.90e-04 & 0.15$\pm$0.03 & 0.09$\pm$0.05 & -0.78$\pm$0.17 & *,2,3 \\
SDSS1610+4724  & 2.508  & 21.15$\pm$0.15 & ...            & 0.69$\pm$0.07  & 0.28$\pm$0.04    & 4.04e-03$\pm$1.76e-03 & 0.23$\pm$0.05 & 0.42$\pm$0.24 & 0.01$\pm$0.18  & *,2,3 \\
SDSS2059-0529  & 2.210  & 20.80$\pm$0.20   & ...            & 0.68$\pm$0.16  & 0.30$\pm$0.04     & 2.26e-03$\pm$1.39e-03 & 0.24$\pm$0.05 & 0.1$\pm$0.08  & -0.26$\pm$0.26 & *,2,3 \\
SDSS2100-0641  & 3.092  & 21.05$\pm$0.15 & ...            & 0.62$\pm$0.07  & 0.27$\pm$0.03    & 2.23e-03$\pm$9.60e-04 & 0.21$\pm$0.04 & 0.18$\pm$0.10  & -0.23$\pm$0.18 & *,2,3 \\
UM673A         & 1.626  & 20.70$\pm$0.10   & ...            & -0.42$\pm$0.15 & 0.03$\pm$0.06    & 1.00e-05$\pm$2.00e-05 & 0.02$\pm$0.05 & ...           & -1.59$\pm$0.20  & *,2,3 \\
eHAQ0111+0641  & 2.027  & 21.50$\pm$0.30   & ...            & 0.44$\pm$0.14  & 0.22$\pm$0.04    & 1.05e-03$\pm$8.40e-04 & 0.17$\pm$0.04 & 0.23$\pm$0.25 & -0.47$\pm$0.34 & *,2,3 \\
\hline
\end{longtable}
\flushleft
{\footnotesize{{\bf References:} Measurements of the DTM, DTG, DTM$_{\rm N}$ and $A_{V, \rm depl}$ are from ($\star$) this work, column densities from (1)~\citet{DeCia2016}; (2)~\citet{Berg2015} and total dust-corrected metallicities from \citet{DeCia2018}.}

\setlength\tabcolsep{2pt}
\begin{longtable}{lllllllllc}
\caption{Milky Way dust properties} 
\label{mwtable}\\
\hline \hline
Star & log N(\ion{H}{I}) & log N(\ion{H}{2}) & [Zn/Fe] & 
DTM & 
DTG& DTM$_{\rm N}$ &
$A_{V, \rm depl}$ &
[M/H]$_{\rm tot}$ & Ref\\
\hline
\noalign{\smallskip}\hline\hline\noalign{\smallskip}
\endfirsthead
\caption{continued}\\
\hline
Star & log N(\ion{H}{I}) & log N(\ion{H}{2}) & [Zn/Fe] & 
DTM & 
DTG& DTM$_{\rm N}$ &
$A_{V, \rm depl}$ &
[M/H]$_{\rm tot}$ & Ref\\
\noalign{\smallskip}\hline\hline\noalign{\smallskip}
\endhead
\hline
\endfoot
\hline
\endlastfoot
1-SCO       & ...           & ...           & 1.53$\pm$0.21 & 0.46$\pm$0.06 & ... & 0.40$\pm$0.08  & ...           & ...            & *,2 \\
23O-ORI     & ...           & ...           & 1.75$\pm$0.08 & 0.56$\pm$0.08 & ... & 0.54$\pm$0.08 & ...           & ...            & *,2 \\
62-TAU      & 20.95$\pm$0.10 & 20.79$\pm$0.10 & 1.56$\pm$0.09 & 0.47$\pm$0.06 & 2.55e-03$\pm$8.30e-04 & 0.41$\pm$0.08 & 0.44$\pm$0.23 & -0.41$\pm$0.13 & *,1 \\
BET-1-SCO   & ...           & ...           & 1.56$\pm$0.23 & 0.52$\pm$0.14 & ... & 0.48$\pm$0.15 & ...           & ...            & *,2 \\
CHI-OPH     & 21.13$\pm$0.10 & 20.63$\pm$0.10 & 1.29$\pm$0.04 & 0.43$\pm$0.05 & 2.80e-03$\pm$7.90e-04 & 0.37$\pm$0.07 & 0.48$\pm$0.24 & -0.33$\pm$0.11 & *,1 \\
DEL-ORI-A   & ...           & ...           & 1.41$\pm$0.08 & 0.48$\pm$0.08 & ... & 0.42$\pm$0.09 & ...           & ...            & *,2 \\
EPS-ORI     & ...           & ...           & 1.61$\pm$0.08 & 0.52$\pm$0.07 & ... & 0.47$\pm$0.09 & ...           & ...            & *,2 \\
EPS-PER     & 20.45$\pm$0.10 & 19.52$\pm$0.10 & 1.38$\pm$0.14 & 0.45$\pm$0.06 & 4.12e-03$\pm$1.33e-03 & 0.39$\pm$0.07 & 0.11$\pm$0.06 & -0.18$\pm$0.13 & *,1 \\
HD-110432   & 20.85$\pm$0.10 & 20.64$\pm$0.10 & 1.51$\pm$0.08 & 0.46$\pm$0.06 & 4.67e-03$\pm$1.72e-03 & 0.40$\pm$0.08  & 0.59$\pm$0.33 & -0.14$\pm$0.15 & *,1 \\
HD-116852   & ...           & ...           & 0.84$\pm$0.04 & 0.34$\pm$0.04 & ... & 0.27$\pm$0.05 & ...           & ...            & *,2 \\
HD-149404   & 21.40$\pm$0.10  & 20.79$\pm$0.10 & 1.41$\pm$0.09 & 0.45$\pm$0.06 & 5.68e-03$\pm$1.37e-03 & 0.39$\pm$0.07 & 1.66$\pm$0.80  & -0.04$\pm$0.09 & *,1 \\
HD-154368   & 21.00$\pm$0.10  & 21.16$\pm$0.10 & 1.29$\pm$0.04 & 0.43$\pm$0.05 & 2.28e-03$\pm$6.90e-04 & 0.37$\pm$0.07 & 0.69$\pm$0.36 & -0.42$\pm$0.12 & *,1 \\
HD-164402   & 21.11$\pm$0.10 & 19.49$\pm$0.10 & 1.15$\pm$0.04 & 0.41$\pm$0.05 & 4.00e-03$\pm$1.13e-03 & 0.34$\pm$0.07 & 0.42$\pm$0.21 & -0.15$\pm$0.11 & *,1 \\
HD-18100    & ...           & ...           & 0.80$\pm$0.06  & 0.33$\pm$0.04 & ... & 0.27$\pm$0.05 & ...           & ...            & *,2 \\
HD-188439   & 20.78$\pm$0.10 & 19.95$\pm$0.10 & 1.20$\pm$0.03  & 0.41$\pm$0.05 & 6.03e-03$\pm$1.22e-03 & 0.35$\pm$0.07 & 0.36$\pm$0.17 & 0.02$\pm$0.07  & *,1 \\
HD-199579   & 21.04$\pm$0.10 & 20.53$\pm$0.10 & 1.65$\pm$0.06 & 0.48$\pm$0.06 & 9.71e-03$\pm$3.37e-03 & 0.42$\pm$0.08 & 1.38$\pm$0.75 & 0.16$\pm$0.14  & *,1 \\
HD-206267   & 21.30$\pm$0.10  & 20.86$\pm$0.10 & 1.42$\pm$0.12 & 0.45$\pm$0.06 & 3.84e-03$\pm$1.16e-03 & 0.39$\pm$0.07 & 1.05$\pm$0.54 & -0.21$\pm$0.12 & *,1 \\
HD-215733   & ...           & ...           & 0.92$\pm$0.13 & 0.35$\pm$0.05 & ... & 0.29$\pm$0.06 & ...           & ...            & *,2 \\
HD-62542    & 20.70$\pm$0.10  & 20.81$\pm$0.10 & 1.50$\pm$0.05  & 0.46$\pm$0.06 & 2.34e-03$\pm$6.60e-04 & 0.40$\pm$0.08  & 0.33$\pm$0.28 & -0.44$\pm$0.11 & *,1 \\
HD-73882    & 21.11$\pm$0.10 & 21.11$\pm$0.10 & 1.26$\pm$0.10  & 0.42$\pm$0.06 & 2.19e-03$\pm$6.30e-04 & 0.36$\pm$0.07 & 0.67$\pm$0.34 & -0.43$\pm$0.11 & *,1 \\
HR-4908     & 21.08$\pm$0.10 & 20.14$\pm$0.10 & 1.40$\pm$0.08  & 0.45$\pm$0.06 & 6.67e-03$\pm$2.02e-03 & 0.39$\pm$0.07 & 0.78$\pm$0.40  & 0.03$\pm$0.12  & *,1 \\
IOT-ORI     & 20.20$\pm$0.10  & 14.69$\pm$0.10 & 1.43$\pm$0.10  & 0.45$\pm$0.06 & 1.200e-02$\pm$3.88e-03 & 0.39$\pm$0.07 & 0.15$\pm$0.08 & 0.28$\pm$0.13  & *,1 \\
KAP-AQL     & 20.90$\pm$0.10  & 20.31$\pm$0.10 & 1.28$\pm$0.05 & 0.43$\pm$0.05 & 3.61e-03$\pm$1.24e-03 & 0.37$\pm$0.07 & 0.34$\pm$0.18 & -0.22$\pm$0.14 & *,1 \\
KSI-PER     & ...           & ...           & 1.52$\pm$0.04 & 0.41$\pm$0.08 & ... & 0.32$\pm$0.10  & ...           & ...            & *,2 \\
LAM-SCO     & ...           & ...           & 1.22$\pm$0.07 & 0.39$\pm$0.09 & ... & 0.33$\pm$0.10  & ...           & ...            & *,2 \\
MU-COL      & ...           & ...           & 1.13$\pm$0.08 & 0.42$\pm$0.05 & ... & 0.36$\pm$0.06 & ...           & ...            & *,2 \\
OMI-PER     & 20.82$\pm$0.10 & 20.60$\pm$0.10  & 1.40$\pm$0.09  & 0.45$\pm$0.06 & 2.54e-03$\pm$1.10e-03 & 0.39$\pm$0.07 & 0.29$\pm$0.17 & -0.39$\pm$0.18 & *,1 \\
PI-SCO      & ...           & ...           & 1.53$\pm$0.10  & 0.47$\pm$0.06 & ... & 0.41$\pm$0.08 & ...           & ...            & *,2 \\
RHO-OPH-A   & 21.63$\pm$0.10 & 20.57$\pm$0.10 & 1.19$\pm$0.07 & 0.41$\pm$0.05 & 9.50e-04$\pm$2.90e-04 & 0.35$\pm$0.07 & 0.37$\pm$0.19 & -0.78$\pm$0.12 & *,1 \\
TAU-CMA     & ...           & ...           & 1.33$\pm$0.10  & 0.50$\pm$0.09  & ... & 0.46$\pm$0.10  & ...           & ...            & *,2 \\
TET-MUS     & 21.15$\pm$0.10 & 19.83$\pm$0.10 & 1.32$\pm$0.04 & 0.43$\pm$0.06 & 8.48e-03$\pm$2.24e-03 & 0.37$\pm$0.07 & 1.03$\pm$0.51 & 0.15$\pm$0.10   & *,1 \\
TET01-ORI-C & 21.54$\pm$0.10 & 17.25$\pm$0.10 & 1.04$\pm$0.05 & 0.38$\pm$0.05 & 1.69e-03$\pm$8.00e-04 & 0.32$\pm$0.06 & 0.44$\pm$0.28 & -0.50$\pm$0.20   & *,1 \\
V600-HER    & ...           & ...           & 0.84$\pm$0.08 & 0.34$\pm$0.04 & ... & 0.28$\pm$0.05 & ...           & ...            & *,2 \\
X-PER       & 20.73$\pm$0.1 & 20.92$\pm$0.10 & 1.33$\pm$0.05 & 0.44$\pm$0.05 & 1.68e-03$\pm$4.10e-04 & 0.37$\pm$0.07 & 0.29$\pm$0.14 & -0.56$\pm$0.09 & *,1 \\
ZET-OPH     & ...           & ...           & 1.65$\pm$0.06 & 0.47$\pm$0.07 & ... & 0.41$\pm$0.09 & ...           & ...            & *,2 \\
ZET-ORI-A   & 20.39$\pm$0.10 & 15.86$\pm$0.10 & 1.41$\pm$0.10  & 0.45$\pm$0.06 & 8.99e-03$\pm$2.17e-03 & 0.39$\pm$0.07 & 0.17$\pm$0.08 & 0.16$\pm$0.09  & *,1 \\
ZET-PUP     & ...           & ...           & 1.21$\pm$0.21 & 0.39$\pm$0.06 & ... & 0.34$\pm$0.07 & ...           & ...            & *,2 \\
\hline
\end{longtable}
\flushleft
{\footnotesize{{\bf References:} Measurements of the DTM, DTG, DTM$_{\rm N}$ and $A_{V, \rm depl}$ are from ($\star$) this work, column densities from (1)~\citet{DeCia2021}; (2)~\citet{Jenkins2009} and total dust-corrected metallicities from \citet{DeCia2021}.

\setlength\tabcolsep{2pt}
\begin{longtable}{lllllllllc}
\caption{LMC dust properties} 
\label{lmctable}\\
\hline \hline
Star & log N(\ion{H}{I})& log N(\ion{H}{2}) & [Zn/Fe] & 
DTM & 
DTG & DTM$_{\rm N}$&
$A_{V, \rm depl}$ &
[M/H]$_{\rm tot}$ & Ref\\
\hline
BI173    & 21.25$\pm$0.05 & 15.64$\pm$0.10 & 0.83$\pm$0.05 & 0.34$\pm$0.04 & 1.54e-03$\pm$3.00e-04 & 0.28$\pm$0.05 & 0.20$\pm$0.05  & -0.49$\pm$0.07 & *,1,2 \\
BI184    & 21.15$\pm$0.04 & 19.65$\pm$0.10 & 0.73$\pm$0.08 & 0.32$\pm$0.04 & 1.86e-03$\pm$4.80e-04 & 0.26$\pm$0.05 & 0.19$\pm$0.06 & -0.38$\pm$0.10  & *,1,2 \\
BI237    & 21.65$\pm$0.03 & 20.05$\pm$0.10 & 0.98$\pm$0.07 & 0.37$\pm$0.05 & 1.65e-03$\pm$2.50e-04 & 0.31$\pm$0.06 & 0.56$\pm$0.12 & -0.50$\pm$0.04  & *,1,2 \\
BI253    & 21.68$\pm$0.03 & 19.76$\pm$0.10 & 0.83$\pm$0.04 & 0.34$\pm$0.04 & 1.39e-03$\pm$1.90e-04 & 0.27$\pm$0.05 & 0.49$\pm$0.10  & -0.53$\pm$0.03 & *,1,2 \\
PGMW3120 & 21.48$\pm$0.03 & 18.30$\pm$0.10  & 0.97$\pm$0.08 & 0.37$\pm$0.05 & 1.64e-03$\pm$4.00e-04 & 0.31$\pm$0.06 & 0.37$\pm$0.11 & -0.50$\pm$0.09  & *,1,2 \\
PGMW3223 & 21.40$\pm$0.06  & 18.69$\pm$0.10 & 1.03$\pm$0.06 & 0.39$\pm$0.05 & 2.71e-03$\pm$7.00e-04 & 0.32$\pm$0.06 & 0.51$\pm$0.17 & -0.30$\pm$0.10   & *,1,2 \\
SK\,6522  & 20.66$\pm$0.03 & 14.93$\pm$0.10 & 0.83$\pm$0.06 & 0.33$\pm$0.04 & 3.36e-03$\pm$6.70e-04 & 0.27$\pm$0.05 & 0.11$\pm$0.03 & -0.14$\pm$0.07 & *,1,2 \\
SK\,66172 & 21.27$\pm$0.03 & 18.21$\pm$0.10 & 1.11$\pm$0.06 & 0.40$\pm$0.05  & 1.91e-03$\pm$4.30e-04 & 0.33$\pm$0.06 & 0.27$\pm$0.08 & -0.46$\pm$0.08 & *,1,2 \\
SK-6619  & 21.87$\pm$0.07 & 20.20$\pm$0.10  & 1.25$\pm$0.13 & 0.42$\pm$0.06 & 2.28e-03$\pm$8.90e-04 & 0.36$\pm$0.07 & 1.31$\pm$0.59 & -0.41$\pm$0.16 & *,1,2 \\
SK\,6635  & 20.85$\pm$0.04 & 19.30$\pm$0.10  & 0.84$\pm$0.14 & 0.34$\pm$0.05 & 4.19e-03$\pm$1.47e-03 & 0.27$\pm$0.06 & 0.22$\pm$0.09 & -0.05$\pm$0.14 & *,1,2 \\
SK\,67101 & 20.20$\pm$0.04  & 14.14$\pm$0.10 & 0.68$\pm$0.13 & 0.31$\pm$0.04 & 5.45e-03$\pm$2.00e-03 & 0.25$\pm$0.05 & 0.06$\pm$0.03 & 0.10$\pm$0.15   & *,1,2 \\
SK\,67105 & 21.26$\pm$0.04 & 19.13$\pm$0.10 & 0.86$\pm$0.07 & 0.21$\pm$0.16 & 5.00e-04$\pm$3.90e-04 & 0.14$\pm$0.17 & 0.05$\pm$0.07 & -0.77$\pm$0.09 & *,1,2 \\
SK-67191 & 20.78$\pm$0.03 & 14.28$\pm$0.10 & 0.52$\pm$0.20  & 0.24$\pm$0.05 & 1.09e-03$\pm$6.20e-04 & 0.19$\pm$0.06 & 0.05$\pm$0.03 & -0.49$\pm$0.23 & *,1,2 \\
SK\,672   & 21.46$\pm$0.12 & 20.95$\pm$0.10 & 1.15$\pm$0.11 & 0.41$\pm$0.05 & 8.40e-04$\pm$1.70e-04 & 0.35$\pm$0.07 & 0.19$\pm$0.07 & -0.83$\pm$0.07 & *,1,2 \\
SK\,67211 & 20.81$\pm$0.04 & 13.98$\pm$0.10 & 0.73$\pm$0.06 & 0.31$\pm$0.04 & 3.77e-03$\pm$5.80e-04 & 0.25$\pm$0.05 & 0.18$\pm$0.04 & -0.06$\pm$0.04 & *,1,2 \\
SK\,675   & 21.04$\pm$0.04 & 19.46$\pm$0.10 & 1.00$\pm$0.05  & 0.37$\pm$0.05 & 1.76e-03$\pm$2.70e-04 & 0.31$\pm$0.06 & 0.15$\pm$0.03 & -0.47$\pm$0.04 & *,1,2 \\
SK\,68129 & 21.62$\pm$0.14 & 20.20$\pm$0.10  & 1.33$\pm$0.14 & 0.44$\pm$0.05 & 4.51e-03$\pm$2.15e-03 & 0.37$\pm$0.07 & 1.47$\pm$0.87 & -0.13$\pm$0.20  & *,1,2 \\
SK\,68135 & 21.48$\pm$0.02 & 19.87$\pm$0.10 & 1.04$\pm$0.07 & 0.39$\pm$0.05 & 1.66e-03$\pm$2.30e-04 & 0.32$\pm$0.06 & 0.38$\pm$0.08 & -0.51$\pm$0.03 & *,1,2 \\
SK\,68140 & 21.51$\pm$0.11 & 20.11$\pm$0.10 & 1.16$\pm$0.10  & 0.41$\pm$0.05 & 4.11e-03$\pm$1.60e-03 & 0.34$\pm$0.07 & 1.03$\pm$0.50  & -0.14$\pm$0.16 & *,1,2 \\
SK\,68155 & 21.47$\pm$0.09 & 19.99$\pm$0.10 & 1.06$\pm$0.05 & 0.39$\pm$0.05 & 4.13e-03$\pm$1.16e-03 & 0.33$\pm$0.06 & 0.93$\pm$0.35 & -0.12$\pm$0.11 & *,1,2 \\
SK\,6826  & 21.65$\pm$0.06 & 20.38$\pm$0.10 & 1.05$\pm$0.15 & 0.38$\pm$0.05 & 9.00e-04$\pm$3.70e-04 & 0.32$\pm$0.07 & 0.31$\pm$0.14 & -0.77$\pm$0.17 & *,1,2 \\
SK\,6852  & 21.31$\pm$0.06 & 19.47$\pm$0.10 & 0.43$\pm$0.12 & 0.21$\pm$0.04 & 1.59e-03$\pm$2.80e-04 & 0.16$\pm$0.04 & 0.23$\pm$0.06 & -0.27$\pm$0.03 & *,1,2 \\
SK\,6873  & 21.68$\pm$0.02 & 20.09$\pm$0.10 & 1.24$\pm$0.12 & 0.38$\pm$0.12 & 3.48e-03$\pm$1.08e-03 & 0.31$\pm$0.13 & 1.25$\pm$0.53 & -0.18$\pm$0.03 & *,1,2 \\
SK\,69104 & 19.57$\pm$0.68 & 14.03$\pm$0.10 & 1.06$\pm$0.27 & 0.39$\pm$0.06 & 1.36e-02$\pm$2.30e-02 & 0.32$\pm$0.07 & 0.04$\pm$0.09 & 0.40$\pm$0.73   & *,1,2 \\
SK\,69175 & 20.64$\pm$0.03 & 14.28$\pm$0.10 & 0.38$\pm$0.12 & 0.20$\pm$0.03  & 8.00e-04$\pm$2.70e-04 & 0.15$\pm$0.04 & 0.02$\pm$0.01 & -0.55$\pm$0.13 & *,1,2 \\
SK\,69246 & 21.48$\pm$0.02 & 19.71$\pm$0.10 & 0.80$\pm$0.04  & 0.33$\pm$0.04 & 1.81e-03$\pm$2.30e-04 & 0.27$\pm$0.05 & 0.40$\pm$0.08  & -0.41$\pm$0.02 & *,1,2 \\
SK\,69279 & 21.63$\pm$0.05 & 20.31$\pm$0.10 & 0.76$\pm$0.06 & 0.32$\pm$0.04 & 7.20e-04$\pm$1.60e-04 & 0.26$\pm$0.05 & 0.22$\pm$0.07 & -0.79$\pm$0.08 & *,1,2 \\
SK\,70115 & 21.18$\pm$0.08 & 19.94$\pm$0.10 & 0.67$\pm$0.05 & 0.30$\pm$0.03  & 2.02e-03$\pm$2.50e-04 & 0.23$\pm$0.04 & 0.22$\pm$0.06 & -0.31$\pm$0.02 & *,1,2 \\
SK\,7079  & 21.34$\pm$0.04 & 20.26$\pm$0.10 & 1.51$\pm$0.09 & 0.39$\pm$0.09 & 4.48e-03$\pm$1.67e-03 & 0.32$\pm$0.10  & 0.74$\pm$0.33 & -0.08$\pm$0.13 & *,1,2 \\
SK\,7145  & 21.11$\pm$0.03 & 18.63$\pm$0.10 & 0.81$\pm$0.06 & 0.34$\pm$0.04 & 2.81e-03$\pm$5.60e-04 & 0.27$\pm$0.05 & 0.27$\pm$0.07 & -0.22$\pm$0.07 & *,1,2 \\
SK\,7150  & 21.24$\pm$0.05 & 20.13$\pm$0.10 & 0.55$\pm$0.05 & 0.28$\pm$0.03 & 1.42e-03$\pm$2.80e-04 & 0.22$\pm$0.04 & 0.17$\pm$0.05 & -0.44$\pm$0.07 & *,1,2 \\
\hline
\end{longtable}
\flushleft
{\footnotesize{{\bf References:} Measurements of the DTM, DTG, DTM$_{\rm N}$ and $A_{V, \rm depl}$ are from ($\star$) this work, column densities from (1)~\citet{Roman-Duval2021}; and total dust-corrected metallicities from (2)~De Cia et al. in prep. }

\setlength\tabcolsep{2pt}
\begin{longtable}{lllllllllc}
\caption{SMC dust properties} 
\label{smctable}\\
\hline \hline
Star & log N(\ion{H}{I})& log N(\ion{H}{2}) & [Zn/Fe] & 
DTM & 
DTG & DTM$_{\rm N}$&
$A_{V, \rm depl}$ &
[M/H]$_{\rm tot}$ & Ref\\
\hline
AzV18  & 22.04$\pm$0.02 & 20.36$\pm$0.08 & 0.91$\pm$0.09 & 0.35$\pm$0.04 & 1.07e-03$\pm$1.60e-04 & 0.29$\pm$0.06 & 0.92$\pm$0.20  & -0.66$\pm$0.03 & *,1,2 \\
AzV26  & 21.70$\pm$0.06  & 20.63$\pm$0.06 & 0.60$\pm$0.05  & 0.28$\pm$0.03 & 4.10e-04$\pm$5.00e-05 & 0.22$\pm$0.04 & 0.17$\pm$0.04 & -0.98$\pm$0.02 & *,1,2 \\
AzV47  & 21.32$\pm$0.04 & 18.54$\pm$0.70  & 0.57$\pm$0.07 & 0.27$\pm$0.03 & 6.90e-04$\pm$1.00e-04 & 0.21$\pm$0.04 & 0.10$\pm$0.02  & -0.73$\pm$0.03 & *,1,2 \\
AzV80  & 21.81$\pm$0.02 & 20.08$\pm$0.30  & 0.69$\pm$0.06 & 0.30$\pm$0.04  & 6.40e-04$\pm$9.00e-05 & 0.24$\pm$0.05 & 0.31$\pm$0.07 & -0.81$\pm$0.02 & *,1,2 \\
AzV95  & 21.49$\pm$0.04 & 19.40$\pm$0.09  & 0.57$\pm$0.07 & 0.27$\pm$0.03 & 5.30e-04$\pm$8.00e-05 & 0.21$\pm$0.04 & 0.12$\pm$0.03 & -0.84$\pm$0.03 & *,1,2 \\
AzV104 & 21.45$\pm$0.06 & 19.23$\pm$0.30  & 0.51$\pm$0.16 & 0.23$\pm$0.04 & 3.90e-04$\pm$1.80e-04 & 0.18$\pm$0.05 & 0.08$\pm$0.04 & -0.92$\pm$0.19 & *,1,2 \\
AzV207 & 21.43$\pm$0.06 & 19.40$\pm$0.10   & 0.83$\pm$0.06 & 0.34$\pm$0.04 & 1.66e-03$\pm$2.50e-04 & 0.27$\pm$0.05 & 0.34$\pm$0.08 & -0.45$\pm$0.04 & *,1,2 \\
AzV216 & 21.64$\pm$0.03 & 18.78$\pm$1.30  & 0.72$\pm$0.08 & 0.30$\pm$0.04  & 7.20e-04$\pm$1.70e-04 & 0.24$\pm$0.05 & 0.23$\pm$0.07 & -0.77$\pm$0.09 & *,1,2 \\
AzV229 & 21.06$\pm$0.04 & 15.66$\pm$0.30  & 0.62$\pm$0.08 & 0.25$\pm$0.03 & 8.30e-04$\pm$1.30e-04 & 0.20$\pm$0.04  & 0.07$\pm$0.02 & -0.62$\pm$0.03 & *,1,2 \\
AzV242 & 21.32$\pm$0.04 & 17.21$\pm$1.30  & 1.16$\pm$0.18 & 0.41$\pm$0.06 & 1.37e-03$\pm$2.30e-04 & 0.35$\pm$0.07 & 0.22$\pm$0.05 & -0.62$\pm$0.04 & *,1,2 \\
AzV321 & 20.70$\pm$0.08  & 14.44$\pm$1.30  & 0.90$\pm$0.11  & 0.35$\pm$0.05 & 2.07e-03$\pm$3.60e-04 & 0.29$\pm$0.06 & 0.08$\pm$0.02 & -0.37$\pm$0.05 & *,1,2 \\
AzV332 & 20.54$\pm$0.16 & 14.50$\pm$0.12  & 0.56$\pm$0.06 & 0.26$\pm$0.03 & 1.20e-03$\pm$5.00e-04 & 0.20$\pm$0.04  & 0.03$\pm$0.02 & -0.47$\pm$0.17 & *,1,2 \\
AzV388 & 21.18$\pm$0.04 & 19.40$\pm$0.10   & 0.99$\pm$0.07 & 0.35$\pm$0.05 & 1.36e-03$\pm$2.60e-04 & 0.30$\pm$0.06  & 0.16$\pm$0.04 & -0.56$\pm$0.06 & *,1,2 \\
AzV456 & 21.00$\pm$0.06  & 20.93$\pm$0.10  & 1.36$\pm$0.09 & 0.44$\pm$0.06 & 5.00e-04$\pm$1.00e-04 & 0.38$\pm$0.07 & 0.11$\pm$0.03 & -1.09$\pm$0.07 & *,1,2 \\
AzV476 & 21.85$\pm$0.07 & 20.95$\pm$0.30  & 0.81$\pm$0.17 & 0.33$\pm$0.04 & 5.20e-04$\pm$2.50e-04 & 0.27$\pm$0.05 & 0.34$\pm$0.18 & -0.95$\pm$0.20  & *,1,2 \\
AzV327 & 20.93$\pm$0.10  & 14.79$\pm$0.10  & 0.93$\pm$0.19 & 0.36$\pm$0.05 & 3.00e-04$\pm$2.00e-04 & 0.30$\pm$0.06  & 0.09$\pm$0.08 & -1.23$\pm$0.29 & *,1,2 \\
AzV238 & 21.41$\pm$0.10  & 15.95$\pm$0.10  & 0.99$\pm$0.04 & 0.38$\pm$0.05 & 1.04e-03$\pm$5.40e-04 & 0.31$\pm$0.06 & 0.20$\pm$0.15  & -0.70$\pm$0.22  & *,1,2 \\
SK116  & 21.57$\pm$0.10  & 18.53$\pm$0.10  & 0.64$\pm$0.04 & 0.29$\pm$0.03 & 5.50e-04$\pm$2.90e-04 & 0.23$\pm$0.04 & 0.15$\pm$0.11 & -0.86$\pm$0.22 & *,1,2 \\
\hline
\end{longtable}
\flushleft
{\footnotesize{{\bf References:} Measurements of the DTM, DTG, DTM$_{\rm N}$ and $A_{V, \rm depl}$ are from ($\star$) this work, column densities from (1)~\citet{Jenkins2017}; and total dust-corrected metallicities from (2)~De Cia et al. in prep. } 

\newpage

\setlength\tabcolsep{2pt}
\begin{longtable}{lllcllllllc}
\caption{GRB-DLAs dust properties} 
\label{grbtable}\\
\hline \hline
GRB & z$_{\rm abs}$ & log N(\ion{H}{I})& log N(\ion{H}{2}) & [Zn/Fe] & 
DTM & 
DTG & DTM$_{\rm N}$&
$A_{V, \rm depl}$ &
[M/H]$_{\rm tot}$& Ref\\
\hline
00926  & 2.0380   & 21.30$\pm$0.20   & ...            & 1.06$\pm$0.18  & 0.39$\pm$0.05 & 8.55e-03$\pm$5.63e-03 & 0.32$\pm$0.07 & 1.30$\pm$1.00   & 0.20$\pm$0.28  & *,1  \\
030226  & 1.9870   & 20.50$\pm$0.30   & ...            & -0.18$\pm$0.12 & ...  & ... & ...  & ...           & -1.07$\pm$0.31 & *,1 \\
050730  & 3.9690   & 22.10$\pm$0.10   & ...            & 0.08$\pm$0.05  & 0.06$\pm$0.02 & 4.00e-06$\pm$2.00e-06 & 0.04$\pm$0.01 & ...           & -2.31$\pm$0.18 & *,1 \\
050820A & 2.6150   & 21.05$\pm$0.10  & ...            & 0.83$\pm$0.05  & 0.34$\pm$0.04 & 1.53e-03$\pm$3.92e-04 & 0.27$\pm$0.05 & 0.13$\pm$0.06 & -0.49$\pm$0.10 & *,1  \\
050922C & 2.1990   & 21.55$\pm$0.10  & ...            & 0.18$\pm$0.46  & 0.12$\pm$0.11 & 1.94e-05$\pm$2.20e-05 & 0.08$\pm$0.10  & 0.005$\pm$0.006  & -1.92$\pm$0.26 & *,1 \\
071031  & 2.6920   & 22.15$\pm$0.05 & ...            & 0.04$\pm$0.03  & 0.03$\pm$0.01 & 7.70e-06$\pm$3.70e-06 & 0.02$\pm$0.01 & 0.007$\pm$0.004  & -1.75$\pm$0.09 & *,1 \\
080413A & 2.4330   & 21.85$\pm$0.15 & ...            & 0.13$\pm$0.07  & 0.09$\pm$0.02 & 3.02e-05$\pm$1.49e-05 & 0.06$\pm$0.02 & 0.01$\pm$0.01 & -1.60$\pm$0.18 & *,1  \\
081008  & 1.9685  & 21.11$\pm$0.10  & ...            & 0.55$\pm$0.04  & 0.26$\pm$0.03 & 1.12e-03$\pm$4.61e-04 & 0.20$\pm$0.04  & 0.10$\pm$0.06  & -0.51$\pm$0.17& *,1  \\
090809A & 2.7373  & 21.48$\pm$0.07 & ...            & 0.75$\pm$0.21  & 0.32$\pm$0.05 & 1.54e-03$\pm$5.85e-04 & 0.26$\pm$0.06 & 0.34$\pm$0.19 & -0.46$\pm$0.15 & *,1 \\
090926A & 2.1069  & 21.58$\pm$0.01 & ...            & 0.88$\pm$0.11  & 0.34$\pm$0.05 & 8.99e-05$\pm$1.59e-05 & 0.28$\pm$0.06 & 0.03$\pm$0.01 & -1.72$\pm$0.05 & *,1 \\
100219A & 4.6676  & 21.28$\pm$0.02 & ...            & 0.12$\pm$0.30   & 0.07$\pm$0.08 & 7.12e-05$\pm$8.05e-05 & 0.05$\pm$0.07 & 0.01$\pm$0.01 & -1.16$\pm$0.11 & *,1 \\
111008A & 4.9910   & 22.39$\pm$0.01 & ...            & 0.22$\pm$0.10   & 0.13$\pm$0.02 & 2.87e-05$\pm$8.60e-06 & 0.09$\pm$0.02 & 0.05$\pm$0.02 & -1.79$\pm$0.10  & *,1 \\
111107A & 2.8930   & 21.10$\pm$0.04  & ...            & 0.70$\pm$0.55   & 0.30$\pm$0.09  & 2.22e-03$\pm$2.40e-03 & 0.24$\pm$0.10  & 0.20$\pm$0.24  & -0.28$\pm$0.45& *,1  \\
120119A & 1.7285  & 22.44$\pm$0.12 & ...            & 0.93$\pm$0.24  & 0.36$\pm$0.06 & 8.16e-04$\pm$7.99e-04 & 0.30$\pm$0.07  & 1.69$\pm$1.78 & -0.79$\pm$0.42& *,1  \\
120327A & 2.8143  & 22.07$\pm$0.01 & 17.39$\pm$0.13 & 0.27$\pm$0.07  & 0.15$\pm$0.03 & 9.78e-05$\pm$1.99e-05 & 0.11$\pm$0.03 & 0.08$\pm$0.03 & -1.34$\pm$0.02 & *,1 \\
120716A & 2.4874  & 21.73$\pm$0.03 & ...            & 0.69$\pm$0.23  & 0.31$\pm$0.05 & 1.15e-03$\pm$2.84e-04 & 0.24$\pm$0.06 & 0.45$\pm$0.21 & -0.57$\pm$0.08 & *,1 \\
120815A & 2.3582  & 22.09$\pm$0.01 & 20.42$\pm$0.08 & 1.01$\pm$0.05  & 0.38$\pm$0.05 & 3.10e-04$\pm$4.30e-05 & 0.31$\pm$0.06 & 0.30$\pm$0.12  & -1.23$\pm$0.03 & *,1 \\
120909A & 3.9290   & 21.82$\pm$0.02 & ...            & 1.15$\pm$0.09  & 0.41$\pm$0.05 & 2.90e-03$\pm$7.62e-04 & 0.34$\pm$0.07 & 1.48$\pm$0.68 & -0.29$\pm$0.10 & *,1  \\
121024A & 2.3005  & 21.78$\pm$0.02 & 19.90$\pm$0.17  & 0.77$\pm$0.08  & 0.31$\pm$0.04 & 8.95e-04$\pm$1.87e-04 & 0.25$\pm$0.05 & 0.41$\pm$0.18 & -0.68$\pm$0.07 & *,1 \\
130408A & 3.7579  & 21.90$\pm$0.01  & ...            & 0.29$\pm$0.07  & 0.17$\pm$0.02 & 7.99e-05$\pm$1.43e-05 & 0.12$\pm$0.03 & 0.04$\pm$0.02 & -1.46$\pm$0.05 & *,1 \\
130606A & 5.9127  & 19.88$\pm$0.01 & ...            & 0.49$\pm$0.10   & 0.24$\pm$0.03 & 8.87e-05$\pm$1.99e-05 & 0.19$\pm$0.04 & ...           & -1.58$\pm$0.08& *,1  \\
140311A & 4.9550   & 22.30$\pm$0.02  & ...            & 0.35$\pm$0.21  & 0.20$\pm$0.05  & 2.76e-05$\pm$9.80e-06 & 0.14$\pm$0.05 & 0.04$\pm$0.02 & -2.00$\pm$0.11  & *,1 \\
141028A & 2.3333  & 20.39$\pm$0.03 & ...            & -0.04$\pm$0.26 & 0.001$\pm$0.090  & 4.00e-07$\pm$2.91e-05 & 0.001$\pm$0.062  & ...           & -1.62$\pm$0.28 & *,1 \\
141109A & 2.9940   & 22.18$\pm$0.02 & 18.02$\pm$0.12 & 0.49$\pm$0.07  & 0.24$\pm$0.03 & 1.45e-04$\pm$2.46e-05 & 0.19$\pm$0.04 & 0.15$\pm$0.06 & -1.37$\pm$0.05& *,1  \\
150403A & 2.0571  & 21.73$\pm$0.02 & 19.90$\pm$0.14  & 0.63$\pm$0.08  & 0.29$\pm$0.04 & 4.84e-04$\pm$8.28e-05 & 0.23$\pm$0.04 & 0.19$\pm$0.08 & -0.92$\pm$0.05& *,1  \\
151021A & 2.3297  & 22.14$\pm$0.03 & 18.99$\pm$1.28 & 0.86$\pm$0.08  & 0.35$\pm$0.05 & 5.20e-04$\pm$1.07e-04 & 0.28$\pm$0.06 & 0.53$\pm$1.02 & -0.97$\pm$0.07 & *,1 \\
151027B & 4.0650   & 20.54$\pm$0.07 & ...            & 0.49$\pm$0.64  & 0.24$\pm$0.12 & 8.67e-04$\pm$6.78e-04 & 0.19$\pm$0.12 & 0.02$\pm$0.02 & -0.59$\pm$0.27 & *,1 \\
160203A & 3.5187  & 21.74$\pm$0.02 & ...            & 0.37$\pm$0.18  & 0.19$\pm$0.04 & 3.25e-04$\pm$7.45e-05 & 0.15$\pm$0.04 & 0.12$\pm$0.06 & -0.92$\pm$0.04 & *,1 \\
161023A & 2.7100    & 20.95$\pm$0.01 & ...            & 0.44$\pm$0.04  & 0.22$\pm$0.02 & 2.76e-04$\pm$3.77e-05 & 0.17$\pm$0.03 & 0.02$\pm$0.01 & -1.05$\pm$0.04 & *,1 \\
170202A & 3.6456  & 21.53$\pm$0.04 & ...            & 0.75$\pm$0.23  & 0.32$\pm$0.05 & 4.24e-04$\pm$1.44e-04 & 0.26$\pm$0.06 & 0.11$\pm$0.05 & -1.02$\pm$0.13 & *,1 \\
181020A & 2.9379  & 22.24$\pm$0.03 & ...            & 0.75$\pm$0.14  & 0.32$\pm$0.04 & 2.80e-04$\pm$6.48e-05 & 0.26$\pm$0.06 & 0.36$\pm$0.16 & -1.20$\pm$0.08 & *,1  \\
190114A & 3.3764  & 22.19$\pm$0.05 & ...            & 1.05$\pm$0.08  & 0.39$\pm$0.05 & 3.64e-04$\pm$6.69e-05 & 0.32$\pm$0.06 & 0.43$\pm$0.18 & -1.17$\pm$0.06& *,1  \\
190106A & 1.8599  & 21.00$\pm$0.04  & ...            & 1.12$\pm$0.10   & 0.40$\pm$0.05  & 2.19e-03$\pm$5.77e-04 & 0.33$\pm$0.06 & 0.17$\pm$0.08 & -0.40$\pm$0.10 & *,1   \\
190919B & 3.2241 & 21.75$\pm$0.06 & ...            & 0.33$\pm$0.33  & 0.18$\pm$0.07 & 1.43e-04$\pm$7.45e-05 & 0.14$\pm$0.07 & 0.05$\pm$0.04 & -1.25$\pm$0.15 & *,1 \\
191011A & 1.7204 & 21.65$\pm$0.08 & ...            & 0.33$\pm$0.09  & 0.18$\pm$0.02 & 5.97e-04$\pm$1.22e-04 & 0.14$\pm$0.03 & 0.18$\pm$0.08 & -0.63$\pm$0.07 & *,1 \\
210905A & 6.3118  & 21.10$\pm$0.10   & ...            & 0.33$\pm$0.09  & 0.18$\pm$0.02 & 4.85e-05$\pm$1.57e-05 & 0.14$\pm$0.03 & ...           & -1.72$\pm$0.13 & *,1 \\
\hline
\end{longtable}
\flushleft
{\footnotesize{{\bf References:} ($\star$) This work; (1)~\citet{Heintz2023}} 

}

\end{document}